\documentclass[review]{elsarticle}

\usepackage{hyperref}
\usepackage{amsmath}
\usepackage{pgfplots}
\usepackage{xcolor}

\journal{International Journal of Heat and Fluid Flow}

\begin{document}
\begin{frontmatter}

\title{A Discrete Immersed Boundary Method for the numerical simulation of heat transfer in compressible flows}


\author[Poitiers]{H. Riahi}
\author[Poitiers]{E. Goncalves da Silva}
\author[Lille]{M. Meldi}
\cortext[Lille]{Corresponding author, \textit{marcello.meldi@ensam.eu}}
\address[Poitiers]{Institut Pprime, Department of Fluid Flow, Heat Transfer and Combustion, CNRS -
ENSMA - Universit\'{e} de Poitiers, UPR 3346, SP2MI - T\'{e}l\'{e}port, 211 Bd. Marie et Pierre Curie,
B.P. 30179 F86962 Futuroscope Chasseneuil Cedex, France}
\address[Lille]{Arts et Métiers ParisTech, CNRS, Univ. Lille, ONERA, Centrale Lille, UMR 9014- LMFL- Laboratoire de Mécanique des fluides de Lille- Kampé de Feriet, F-59000 Lille, France}

\begin{abstract}
In the present study, a discrete forcing Immersed Boundary Method (IBM) is proposed for the numerical simulation of high-speed flow problems including heat exchange. The flow field is governed by the compressible Navier-Stokes equations, which are resolved by using the open source library OpenFOAM. The numerical solver is modified to include source terms in the momentum equation and in the energy equation, which account for the presence of the immersed body. The method is validated on some benchmark test cases dealing with forced convection problems and moving immersed bodies. The results obtained are in very good agreement with data provided in the literature. The method is further assessed by investigating three-dimensional high Mach flows around a heated sphere with different wall temperature. Even for this more complex test case, the method provides an accurate representation of both thermal and velocity fields.
\end{abstract}

\begin{keyword}
immersed boundary method, heat transfer, forced convection, compressible flows,
fluid-structure interaction, OpenFOAM.
\end{keyword}

\end{frontmatter}


\section{Introduction}

With the rapid increase of computational resources available in
computational centers, the numerical representation of flows around
complex configurations of industrial interest is becoming an established
reality. A large effort by the scientific community in the last
decades has allowed to identify strategies for the numerical
investigation of such problems. As a result, new methods studied to
solve problematic aspects of the simulations are getting to maturity. Among
these, the Immersed Boundary Method (IBM)
\cite{Peskin1972,Peskin2002,Mittal_2005,Sotiropoulos_2014,Kim2019_ijhff} is an increasing popular
tool for the representation of moving bodies. In the IBM, the
presence of the immersed body is represented via the inclusion of
source terms in the dynamic equations of the physical model. The
usage of the IBM allows to naturally exclude problematic aspects of
the mesh procedure, such as deformed / stretched mesh elements for
the representation of complex geometries. These aspects can result
in very large numerical errors, significantly affecting the
prediction of the flow features. In addition, IBM can naturally
account for the immersed body motion / deformation with almost zero
computational resources, excluding expensive re-meshing procedures.

While many different IBM proposals can be found in the literature
for purely dynamic effects associated with the immersed body
\cite{Beyer1992,Goldstein_jcp_1993,Fadlun2000,uhlmann_jcp_2005,Cheny2010,Kumar_2020}
(i.e. focusing on the determination of a source forcing term for the
momentum equation) the analysis of heat transfer via IBM is less
investigated and mainly dealing with incompressible flow simulation
\cite{Kim2001,Ren2013,Park2017}.  {Recent works on the subject have
been proposed by Luo et al. \cite{Luo2015,Luo2016} on fixed bodies. Also, simulations of compressible flows in interaction with moving solids are still rare (see for example Qu et al. \cite{Qu_2018} and Khalili et al. \cite{Khalili_2018}) and very few contributions can be found including the coupling with heat exchange effects \cite{Luo2016}.}	 Accounting for
heat transfer effects in IBM applications for compressible flows
will provide significant advances for a number of relevant high
speed applications, such as the flow in aerospace engines. In fact,
many features of the flow such as the behavior of the boundary layer
of coherent structure organization in turbulent wakes is affected by
thermal phenomena occurring between the flow and the surface of the
immersed body. The correct representation of such phenomena is
essential for an accurate prediction of heat dissipation in complex
mechanical systems.

In the present work, the IBM tool recently developed for integration in the opensource platform \textit{OpenFOAM} \cite{Constant_2017,Riahi_2018} is extended to the analysis of heat exchange for compressible flows. To do so, an explicit term is included in the energy equation of the dynamic system. The article is structured as follows. In section \ref{sec:numIBM} numerical details about the dynamic system and the IBM method are provided and the implementation of the IBM method within the numerical solver is detailed. In section \ref{sec::Validation} the research tool is validated with the analysis of academic test cases. In section \ref{sec::Sphere}, a three-dimensional application is analyzed, considering the flow around a heated sphere. In section \ref{sec:FSI}, elements of fluid-structure interaction are discussed via the analysis of an oscillating cylinder. Finally, concluding remarks are drawn in section \ref{sec:conclusion}.

\section{Numerical ingredients and Immersed Boundary Method}\label{sec:numIBM}

The IBM algorithm developed for heat transfer prediction is here
introduced and its analytic derivation is described.
\subsection{Governing equations}
The starting point is represented by the general Navier--Stokes
equations for a compressible fluid:
\begin{eqnarray}
\label{eq:Mass}
\frac{\partial \rho}{\partial t} &+& div(\rho \mathbf{u}) =0 \\
\label{eq:Momentum}
\dfrac{\partial\rho \mathbf{u}}{\partial t} &+& \mathbf{div}(\rho \mathbf{u} \otimes \mathbf{u}) = - \mathbf{grad} p + \mathbf{div} \overline{\overline{\tau}} + \mathbf{f} \\
\label{eq:Energy}
\dfrac{\partial\rho (e +\mathbf{u} \cdot \mathbf{u}/2)}{\partial t} &+& div (\rho \,(e+\mathbf{u} \cdot \mathbf{u}/2) \, \mathbf{u}) = \nonumber \\
 &-& div( p \, \mathbf{u}) + div (\overline{\overline{\tau}} \mathbf{u}) + div (\lambda(\theta) \mathbf{grad} \theta)+ \mathbf{f} \cdot \mathbf{u} + q
\end{eqnarray}
where $\rho$ is the density, $p$ is the pressure, $\mathbf{u}$ is the velocity, $\overline{\overline{\tau}}$ is the tensor of the viscous constraints, $e$ is the internal energy, $\lambda$ is the thermal conductivity, $\theta$ is the temperature and $\mathbf{f}$ is a general force term.
In the case Newtonian fluids are considered, the tensor $\overline{\overline{\tau}}$ can be written as:

\begin{eqnarray}
\overline{\overline{\tau}}=\mu(\theta) \left( (\overline{grad}\mathbf{u}+^{t}\overline{grad}\mathbf{u})- \frac{2}{3} div(\mathbf{u}) \right)
\end{eqnarray}
where $\mu$ is the dynamic viscosity. Its value is calculated using
the Sutherland's law as a function of temperature $\theta$. $q$ is a
general source term for the energy equation. This system is closed using the equation
of state for perfect gas $p=\rho r \theta$ where $r$ is the specific
gas constant.

\subsection{Immersed Boundary Method for compressible flows: numerical formulation}
\label{sec:IBMcompressible} The IBM proposed in this work roots in
recent proposals by Riahi et al. \cite{Riahi_2018} which was
successfully used for the analysis of compressible flows around
immersed bodies with adiabatic walls. The IBM strategy relies on
communication between an Eulerian mesh used for calculation of the
flow and Lagrangian markers representing the discretized shape of
the immersed geometry. The communication between the two frames of reference allows for the calculation of the dynamic effects acting on the body surface in the Lagrangian space and for their representation on the Eulerian mesh as source terms. 

The first step is the interpolation of physical quantities calculated on the Eulerian mesh to the Lagrangian markers. Lowcase type is used for information on the Eulerian setting, while capital letters (or via the subscript $L$ for Greek letters) are used to indicate physical quantities on the Lagrangian markers. The interpolation operator $\mathcal{I}$ used in this work is:

\begin{equation}
\label{eq:interp}
\mathcal{I}[\mathbf{p}]_{X_s} = [\mathbf{P}](X_s) = \sum_{j\in D_s} (\mathbf{p})_{j}^n \delta_h(\mathbf{x}_{j} - \mathbf{X}_s) \Delta \mathbf{x}
\end{equation}

where $\mathbf{p}$ and $\mathbf{P}$ are a physical quantity, scalar or vectorial ($\theta$ or $\rho \mathbf{u}$ for example) and $\mathbf{x}$ and $\mathbf{X}$ represent the physical coordinates. For each Lagrangian marker $s$,  $D_s$ represents the set of points of the Eulerian mesh from which information is extracted (computational stencil) and the interpolation kernel $\delta_h$ is the discretized delta function used in Pinelli et al. \cite{Pinelli_JCP_2010}:
 {
\begin{equation}
\delta_h(r) = \begin{cases} 
\frac{1}{3} \left( 1 + \sqrt{1 - 3 r^2} \right) \quad 0 \leq r \leq 0.5 \\
\frac{1}{6} \left(5 - 3r - \sqrt{1 - 3 (1 -r)^2} \right) \quad 0.5 \leq r \leq 1.5 \\
0 \quad \text{otherwise}
\end{cases}
\end{equation}
}

 { $\Delta \mathbf{x}$ refers to an Eulerian quadrature, which is $\Delta \mathbf{x}=\Delta x \Delta y \Delta z $ for a Cartesian mesh with uniform distribution.} This step is essential for the calculation of the dynamic effects on the immersed body surface. These effects, which will be detailed in section \ref{sec:forcage_energie1}, are represented as source terms for both the momentum equation ($\mathbf{F}$) and the energy equation ($Q$).

The second step of the IBM procedure is the spreading of the volume force calculated in the first step to the Eulerian mesh elements. We will refer to this term as $\mathbf{R}$, knowing that the numerical procedure for $\mathbf{F}$ and $Q$ is the same. The value of the source term evaluated on the Eulerian mesh, $\mathbf{r}(\mathbf{x}_j)$, is given by:
\begin{equation}
\label{eq:force2}
\mathbf{r}(\mathbf{x}_j) = \sum_{k \in D_j} \mathbf{R}_k \delta_h(\mathbf{x}_{j} - \mathbf{X}_k) \boldsymbol{\epsilon}_k
\end{equation}
 {The $k$-index controls a loop over the Lagrangian markers whose support contains the Eulerian node $j$.} $\boldsymbol{\epsilon}_k$ is the Lagrangian quadrature, which is calculated by solving a linear system to satisfy a partition of unity condition.  {This system can be written in the form}
 {
\begin{equation}
A \mathbf{\epsilon} = \mathbf{1}
\end{equation}
}

 {where $\mathbf{\epsilon} = (\epsilon_1, \cdots, \epsilon_{N_s})^t$ $\mathbf{1} =(1, \cdots, 1)^t$ are vectors whose size is equal to the number of Lagrangian markers $N_s$.
On the other hand, the component $A_{kl}$ of the matrix $A$ is the product between the $k^{th}$ and the $l^{th}$ interpolation kernels:}

 {
\begin{equation}
A_{kl} = \sum_{j \in D_l} \delta_h(x_j-X_k) \, \delta_h(x_j-X_l)
\end{equation}
}

The calculation of the dynamic effects associated with the presence of the immersed body in step one is thus crucial for obtaining a precise representation of the flow. The research team \cite{Riahi_2018} was able to derive an expression for the source term of the momentum equation $\mathbf{F}$ which takes into account the iterative nature of numerical solvers used for the calculation:

\begin{equation}
\label{eq:forceMomentum}
\mathbf{F} = a_s \left( (\rho_L \mathbf{U})^{d} - \overline{(\rho_L \mathbf{U})} \right) 
\end{equation}

where $a_s$ is a coefficient derived by the discretization procedure.  
The superscript $d$ over the physical quantities represents the target behavior of the flow (i.e. zero velocity at the wall) while the overbar represents the result of the interpolation of the quantities from the Eulerian mesh performed in step one. 

The novelty of the present approach is represented by the inclusion
of a volume source term $q$ for the energy equation, which mimics
heat transfer phenomena on the body surface. These effects are
derived from the discretized set of equations via analytic
manipulation. In the present analysis, the heat transfer will be
considered equivalent to a boundary condition applied on the surface
of the immersed body. The analysis will be restricted to the case of
\textit{imposed temperature at the wall}, which is equivalent to
imposing a Dirichlet condition for the temperature field.

\subsubsection{Imposed temperature on the body surface - analytic derivation}
\label{sec:forcage_energie1} 
Similarly to previous works
\cite{uhlmann_jcp_2005,Riahi_2018}, the starting point is
represented by the discretized set of equation in the Eulerian frame
of reference for the mesh element of coordinate $x$. Time advancement from the
instant $n$ to $n+1$ is considered. For sake of simplicity, the procedure for the momentum equation and the forcing $\mathbf{F}$ (detailed in \cite{Riahi_2018}) is not reported here, and the development is provided only for the internal energy equation \ref{eq:Energy}. Its discretized form for an iterative solver can be written as:
\begin{equation}
\label{eq:corr3.1}
a_e \, (\rho e)^{n+1} = {\phi}_e^{n+1/2}  + f_e^{n+1/2} + q_e^{n+1/2} 
\end{equation}
with:
\begin{itemize}
\item $a_e$ : coefficient derived by the discretization procedure
\item ${\phi}_e$ : discretization term calculated at the intermediate
time $n+\dfrac{1}{2}$ which includes the contribution of the
convection term, the heat flux, the pressure work and the viscous
friction work.
\item $ f_e^{n+1/2} $ : discretized form of the term $\mathbf{f} \cdot \mathbf{u}$
\item $ q_e^{n+1/2} $ : discretization of the volume source term $q$. This term represents the heat exchange
between the flow and the immersed body, calculated at the
intermediate time $n+\dfrac{1}{2}$
\end{itemize}

One must keep in mind that $\mathbf{f}$ represent the discretized
volume source term representing the effect of the body in the
momentum equation, as in classical IBM strategies. Thus, the term
$q_e$ is its correspondent in the energy equation, which describes
the heat transfer between the flow and the immersed body. Thus, if
one targets a behavior of the internal energy $(\rho e)^{d}$ at the
instant $n+1$, then the optimized value of the volume source term
$q_e$ must comply with the equation:
\begin{equation}
\label{eq:corr3.3}
q_e^{n+1/2} =a_e \, (\rho e)^{d} - {\phi}_e^{n+1/2} - f_e^{n+1/2}
\end{equation}

By applying the Eulerian-Lagrangian transformation via the
interpolation operator proposed by Pinelli et
al.\cite{Pinelli_JCP_2010}, the equation  \ref{eq:corr3.3} on the Lagrangian markers representing the body surface in the IBM method is
transformed in:
\begin{equation}
\label{eq:corr3.4}
{Q}_e^{n+1/2}=a_e \, (\rho_L E)^{d} - {\Phi}_e^{n+1/2} - F_e^{n+1/2}
\end{equation}
\begin{itemize}
\item $(\rho_L E)^d$ : target internal energy value on the Lagrangian markers
\item ${\Phi}_e^{n+1/2}$ : total contribution of the convection term, the pressure
work term, the viscous friction work term and heat flux term
interpolated on the Lagrangian marker at the time $n+\dfrac{1}{2}$
\end{itemize}

Let us now consider again equation \ref{eq:corr3.1} in the case of the
presence of the body surface does not introduce direct thermal interactions with the flow. This state will be referred to as neutral behavior. In this case, the time advancement to $n+1$ (a suffix $l$ is here used) reads as:
\begin{equation}
\label{eq:corr3.5}
a_e \, (\rho e)^{l} = {\phi}_e^{n+1/2} +f_e^{n+1/2}
\end{equation}

If one further performs a transformation from the Eulerian system to
the Lagrangian frame of reference, equation \ref{eq:corr3.5}
becomes:
\begin{equation}
\label{eq:corr3.6}
a_e \, (\rho_L E)^{l} = {\Phi}_e^{n+1/2} + F_e^{n+1/2} 
\end{equation}
where $(\rho_L E)^{l}$ represents the internal energy at a
Lagrangian point $X_{s}$.

Combining the two equations \ref{eq:corr3.4} et \ref{eq:corr3.6}, an
analytic expression for the volume source term for the energy
equation to be included in the IBM method is derived:
\begin{equation}
\label{eq:corr3.7}
{Q}_e^{n+1/2}=a_e \, ((\rho_L E)^{d} - (\rho_L E)^{l})
\end{equation}
Equation \ref{eq:corr3.7} can be further manipulated if one
considers that $e= c_{v} \theta$ for a perfect gas. Therefore, the
volume source term for the internal energy equation can be written
as a function of the temperature:
 \begin{equation}
\label{eq:corr3.8}
{Q}_e^{n+1/2}=a_e \, c_{v} ((\rho_L \Theta)^{d} - (\rho_L \Theta)^{l})
\end{equation}
with
\begin{itemize}
\item $c_{v}$ : the thermal capacity at constant volume of the interpolated fluid at Lagrangian markers
\item $(\rho_L \Theta)^{d}$ : the desired temperature density at the Lagrangian markers
\item $(\rho_L \Theta)^{l}$ : interpolated temperature density at the Lagrangian markers for
a neutral system.
\end{itemize}

A choice relying on numerical arguments has also been performed for the term $\mathbf{f} \cdot \mathbf{u}$ in the energy equation, which has been systematically set to zero. This term is naturally zero when the body is still, but it shows non-zero values in the case of body movement. However, this term is always very small so interpolation errors are most of the time larger in magnitude, leading to numerical instability and degradation of the accuracy. We therefore decided to remove it from the energy equation.

\subsubsection{IBM implementation in OpenFOAM numerical solvers}\label{sec3:impl}

The implementation of the IBM algorithm has been performed in the
open source library OpenFOAM, owing to previous development of the
research group \cite{Constant_2017,Riahi_2018}. Two solvers for
compressible flows have been considered: \textit{sonicFoam} and
\textit{rhoCentralFoam}. These two solvers are studied to provide
optimal performance for different ranges of $Ma$ numbers. The main
features of the solvers are discussed in \cite{Riahi_2018}. The
newly generated solvers will be referred to in the following as
\textit{IBM-HT} versions of the initial solver modified and are now
presented. One must consider that the versions of the solvers
proposed in \cite{Riahi_2018} are \textit{neutral} with respect to
the heat transfer behavior i.e. the presence of the body surface is
transparent with respect to heat exchange. Details about the algorithmic structure of \textit{IBM-HT-sonicFoam} and \textit{IBM-HT-rhoCentralFoam} are provided in \ref{apendixA} and \ref{apendixB}, respectively.

 {For every test case investigated with these solvers, the numerical discretization used for the gradient and divergence operators in the dynamic equations is linear (i.e. second order centered schemes). A van Leer limiter, which improves the flow prediction in the regions where shocks are observed, is used for the flux interpolation. Also, a grid convergence study is provided in \ref{sec:appC}}

\section{Numerical validation of the \textit{IBM-HT} algorithms} \label{sec::Validation}

Validation of the new solvers is performed on the 2D flow around a
heated circular cylinder. This classical test case has been
extensively investigated in the literature for a large spectrum of
values of $Re$, $Ma$ and temperature ratio
$r_\theta=\frac{\theta_{wall}}{\theta_{inlet}}$, where $\theta_{wall}$ is the wall temperature of the immersed body and $\theta_{inlet}$ is the temperature imposed at the inlet. Numerous
databases are available for comparison.

\subsection{Test case - numerical details}
\label{sec::validationCylinder}
\begin{figure}[ht]
\centering
\includegraphics [width=0.9\linewidth, height=0.5\linewidth]{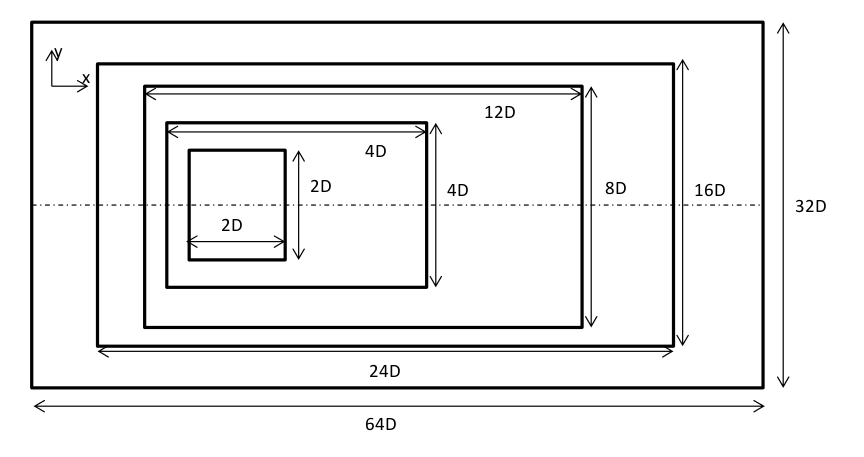}
\caption{\label{fig::fig_thermique} 2D computational domain used for \textit{IBM-HT} validation.}
\end{figure}

The center of the cylinder is the origin of the domain in $(0,0)$ and its diameter is equal to $D$.
The dimensions of the computational domain are $[-16D,48D] \times
[-16D,16D]$  in the streamwise direction $(x)$ and normal direction
$(y)$, respectively. The mesh elements are uniformly distributed in
the central zone near to cylinder $x \times y \in [-D, \,D] \times
[-D, \, D]$. Their size is equal to $\Delta x = \Delta y = 0.01D$.
Outside this region, the mesh is progressively coarsened moving
through five different levels as shown in Figure
\ref{fig::fig_thermique}. The passage from one level to another
implies a coarsening ratio of the elements of $2$ in both $x$ and
$y$ directions. This value is smoothed in correspondence of the
boundaries of the refinement regions.  {This transition in the mesh is generated via an OpenFOAM native tool, \textit{snappyHexMesh}, and the Finite Volume discretization used in the solvers suitably takens it into account. In fact, the solver imposes conservation of the fluxes through the faces of the mesh elements, granting conservativity of the Navier--Stokes equation even with abrupt changes in the resolution of the mesh. Nonetheless, we have paid attention that these transition regions do not occur near the Eulerian supports for each Lagrangian Marker, for this test case and all the other flow configurations presented in the following Sections. This tasks has been easy to perform as it is connected with the need to impose a refined mesh distribution in the vicinity of the immersed body.} The total number of mesh
elements that result is equal to $1.5 \times 10^5$.  {The positioning of the Lagrangian markers is performed so that each Eulerian mesh element crossed by the body surface includes at least one marker. This strategy has been performed as the original version of the method has shown sensitivity to \textit{holes} in the connection between the Eulerian and the Lagrangian space \cite{Pinelli_JCP_2010}. The IBM library integrates an internal check routine which keeps only the Lagrangian marker closer to the center of each mesh element, if one or more markers are detected. For the mesh previously introduced, $401$ Lagrangian markers have been used.}

 {The boundary conditions have been chosen accordingly to general guidelines in the literature and using the team's experience for each of the $Ma$ number investigated, selecting different options for subsonic and supersonic flows. More precisely, a} constant
velocity inlet condition is imposed upstream (left side), a mass
conserving outlet condition is prescribed downstream and  {slip (subsonic) or non
reflective (supersonic)} conditions are imposed in the normal direction.

For each case analyzed, a number of different coefficients are
compared with available data of the literature. Aerodynamic forces
are determined via the bulk flow coefficients known as drag
coefficient $C_d$ and lift coefficient $C_l$:

\begin{equation}
\label{eq:Bulk}
C_d = \frac{2 F_x}{\rho_{\infty} U_{\infty}^2} \; , \qquad C_l = \frac{2 F_y}{\rho_{\infty} U_{\infty}^2}
\end{equation}

the forces in the streamwise direction $F_x$ and in the normal $F_y$
are directly calculated on the Lagrangian markers. $\rho_{\infty}$
and $U_{\infty}$ indicate asymptotic physical quantities imposed at
the inlet.

\subsection{Unsteady flow around a heated 2D circular cylinder}
The flow around a heated circular cylinder in unsteady regime is now
investigated to further assess the performance of the source term
for the internal energy equation. The non-dimensional parameters are
set to $Ma=0.01$ using temperature ratio $r_\theta = 1.8$ and three
Reynolds number $Re= \{100 , 140 , 260\}$. For these cases,
variations in the temperature field significantly influence the flow
organization, especially when the temperature ratio $r_\theta$ is
larger than 1.1 \cite{depalma_2006}.

\begin{figure}
\begin{tabular}{cc}
\includegraphics[width=0.48\linewidth]{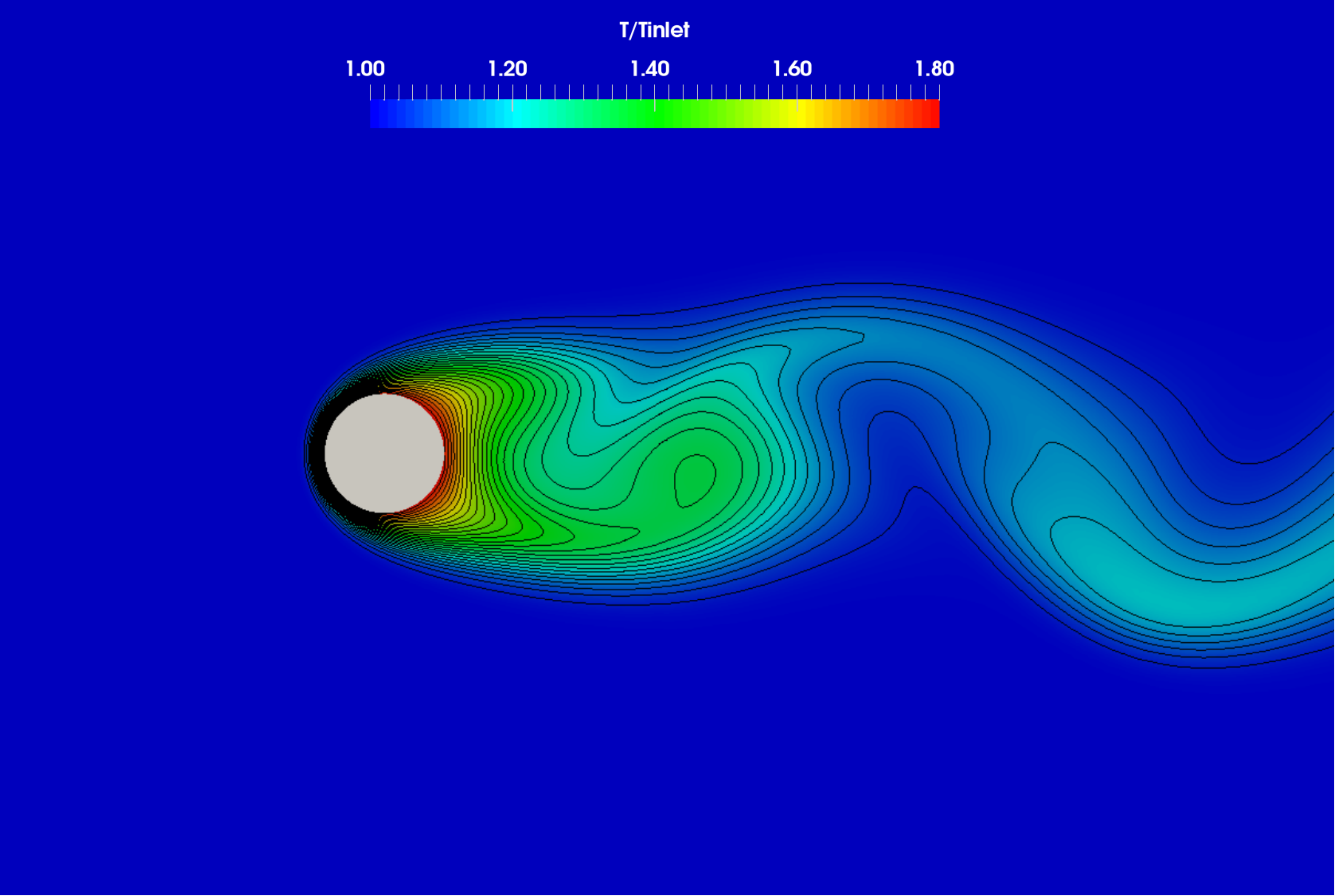} & \includegraphics[width=0.48\linewidth]{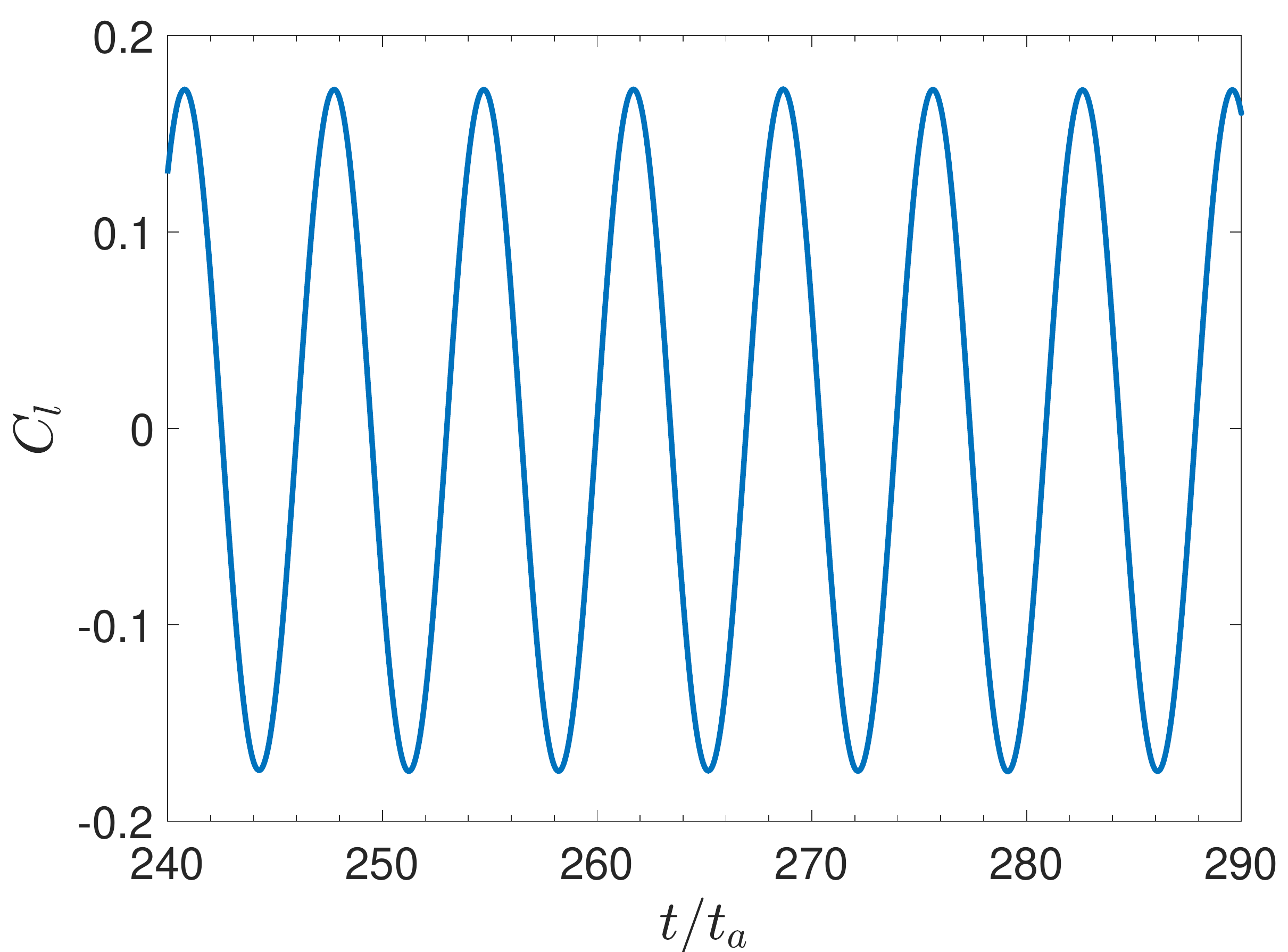} \\
(a) & (b)
\end{tabular}
\caption{\label{fig:instationnaire} Isocontours of the temperature
(a) and time evolution of the lift coefficient (b) for the flow past
a circular cylinder for $Ma=0.01$, $Re = 100$ and $r_\theta=1.8$.
$t_a = D / U_{\infty}$ is the characteristic advection time.}
\end{figure}

For the different cases, an unsteady behavior characterized by a
periodic von K\'arm\'an wake is observed. Qualitative visualizations
of the instantaneous regimes are shown with the temperature
isocontours in Figure \ref{fig:instationnaire} $(a)$. In addition
the time evolution of the lift coefficient is shown in Figure
\ref{fig:instationnaire} $(b)$.

\begin{figure}
\center \includegraphics[width=0.8\linewidth]{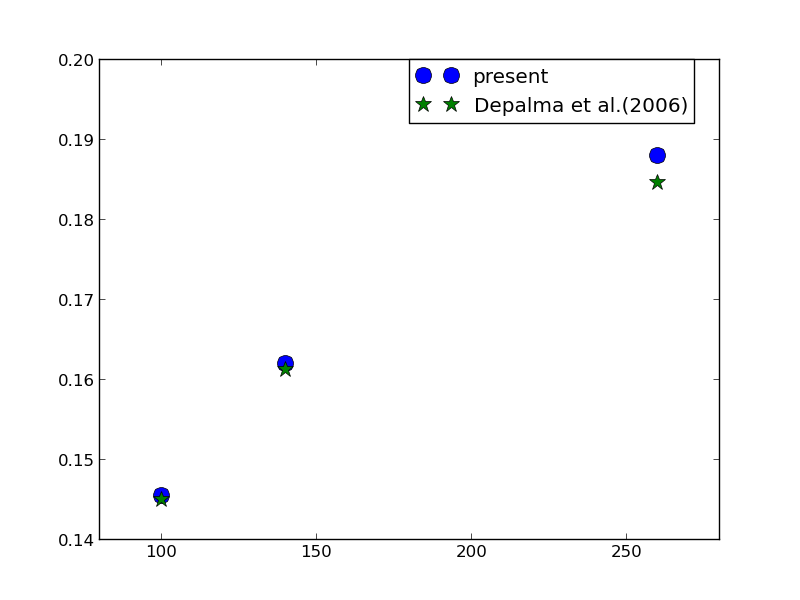}
\caption{\label{ta:st_2d} Strouhal number versus Reynolds number for flow past a heated cylinder: comparison between present results and literature data}
\end{figure}

One important parameter describing the flow is the Strouhal number
$S_t=\frac{f D}{U_\infty}$, where $f$ is the shedding frequency
computed using the time evolution of the lift coefficient $C_l$.
Comparison shows a good agreement with results available in the
literature \cite{depalma_2006,deTullio_2007}, see Figure \ref{ta:st_2d}.
We observe that the vortex shedding frequency, $f$, and thus the Strouhal
number $S_t$, increases for increasing values of $Re$ for a given
$r_\theta$.
Finally, in table \ref{ta:table3} the computed bulk flow
quantities for the flow for all values of $Re$ are reported.  {It can
be seen that the average value of the drag coefficient $C_d$ decreases when $Re$ increases. On the other
hand, the maximum value of the coefficient of lift $C_l^{max}$ exhibits the opposite variation, increasing with $Re$. These results are in agreement with the findings by Homsi et al. \cite{Homsi_2021}.} 
\begin{table}[h!]
\centering
\begin{tabular}{c|c|c}
Present study - $\mathbf{r}_\theta=1.8$ & $C_d$  &  $C_l^{max}$  \\ \hline
$Re=100$ & 1.48    & 0.18 \\
$Re=140$ & 1.46    & 0.29 \\
$Re=260$ & 1.45    & 0.53 \\
\end{tabular}
\caption{ $C_d$ and $C_l^{max}$ for unsteady flow around heated
circular cylinder at $Ma=0.01$. } \label{ta:table3}
\end{table}

\section{Forced convection from an heated sphere in a steady regime} \label{sec::Sphere}

The three-dimensional flow around a sphere is now investigated. This test case, which is significantly more complex, allows for a robust validation of the capability of the IBM method to accurately capture heat exchange.

\subsection{Numerical and computational ingredients}

\begin{figure}[ht]
\centering
\includegraphics [width=0.9\linewidth, height=0.5\linewidth]{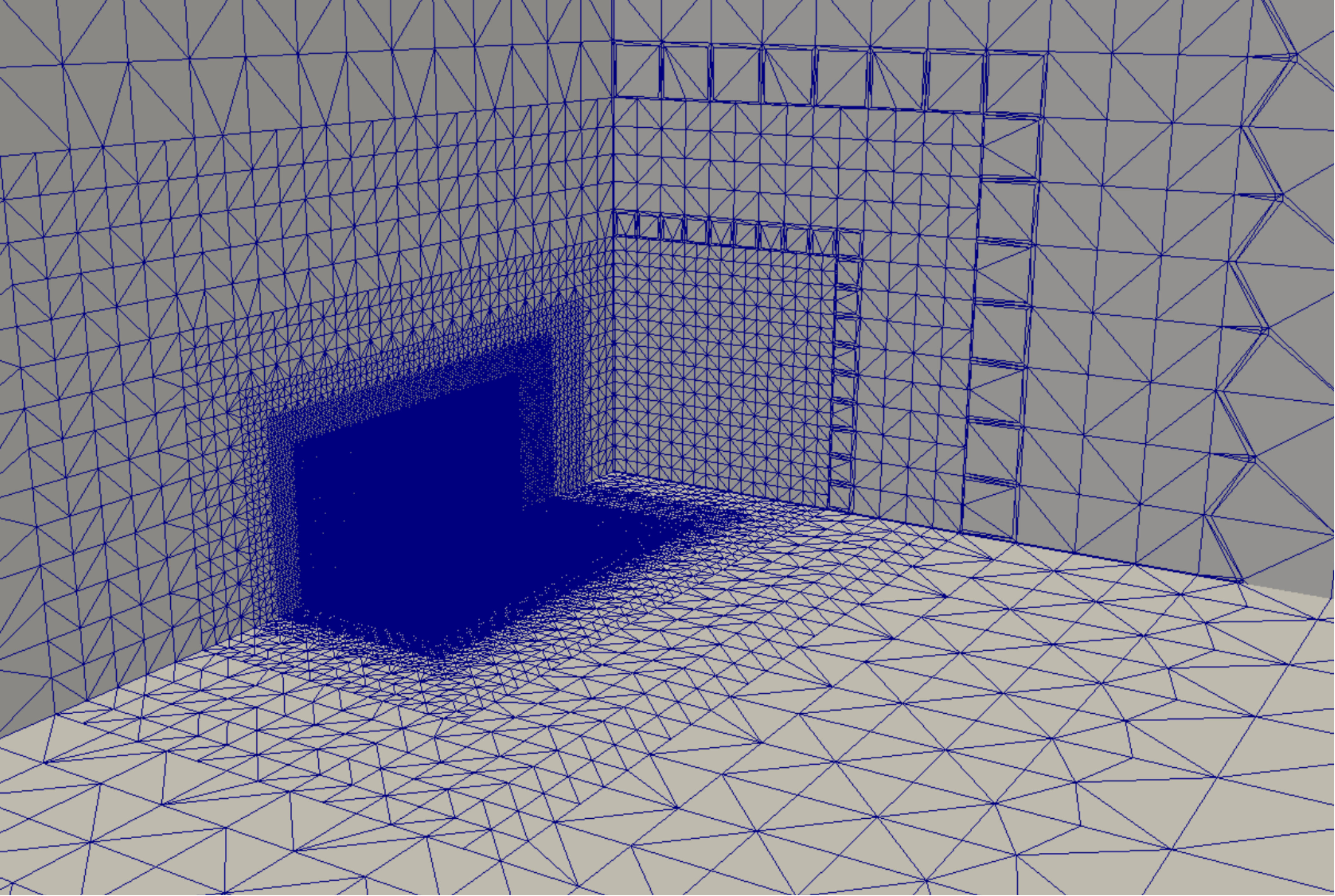}
\caption{\label{fig::fig7} Visualization of cutting planes inside the 3D mesh used for the calculation of the flow around a sphere.}
\end{figure}

The computational domain is here set to $x \times y \times z =
[-16D, \, 48D] \times [-16D, \, 16D] \times [-16D, \, 16D]$ where
$D$ is the diameter of the sphere. Again, the center of the body is
set in the origin of the system and the mesh is obtained using the tool \textit{snappyHexMesh}. The mesh is made by hexahedral
uniform elements which are progressively refined approaching the
sphere region (see figure \ref{fig::fig7}). The size of the elements
is refined by a factor two in each space direction when crossing the
prescribed interfaces between regions at different resolution. The
central most refined region is defined by the coordinates $x \times
y \times z = [-1.25D, \, 1.25D] \times [-1.25D, \, 1.25D] \times
[-1.25D, \, 1.25D]$. Within this region, the mesh resolution is
$\Delta x=\Delta y=\Delta z = 1/64 D$. This mesh is composed by a
total of $5 \times 10^6$ elements and was already used for analysis of compressible flows \cite{Riahi_2018}.
 {The choice of the number and positioning of the Lagrangian markers follows the same strategy presented for the flow around a circular cylinder discussed in sec. \ref{sec::validationCylinder}. Here $12119$ Lagrangian markers are used.}

In order to provide a suitable representation of the physical
features of the flow, the \textit{IBM-HT-rhoCentralFoam} solver is
employed for this study.

\subsection{Physical regimes observed for several $r_\theta$}

\begin{figure}
\begin{tabular}{cc}
\includegraphics[width=0.48\linewidth]{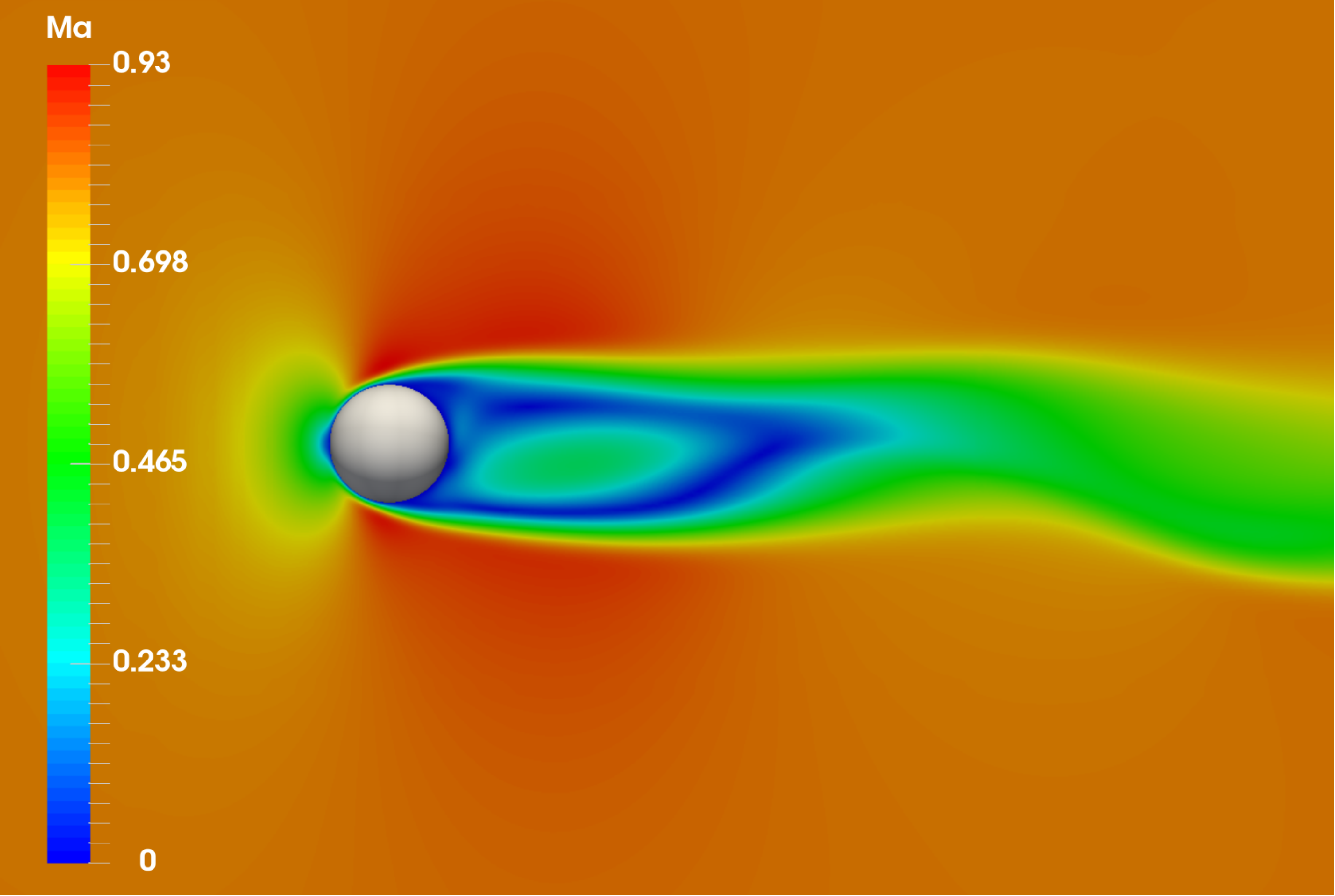} & \includegraphics[width=0.48\linewidth]{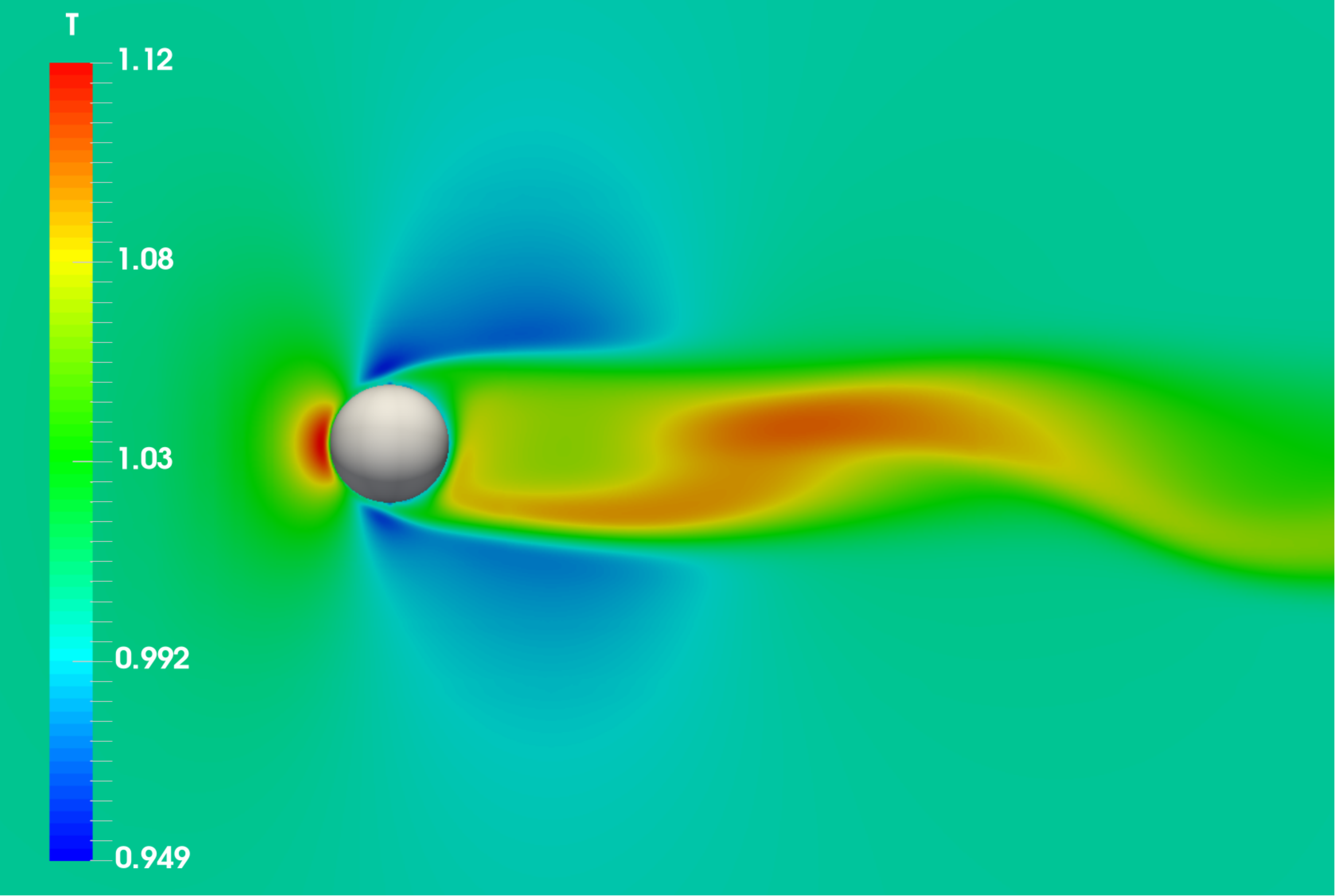} \\
(a) & (b) \\
\includegraphics[width=0.48\linewidth]{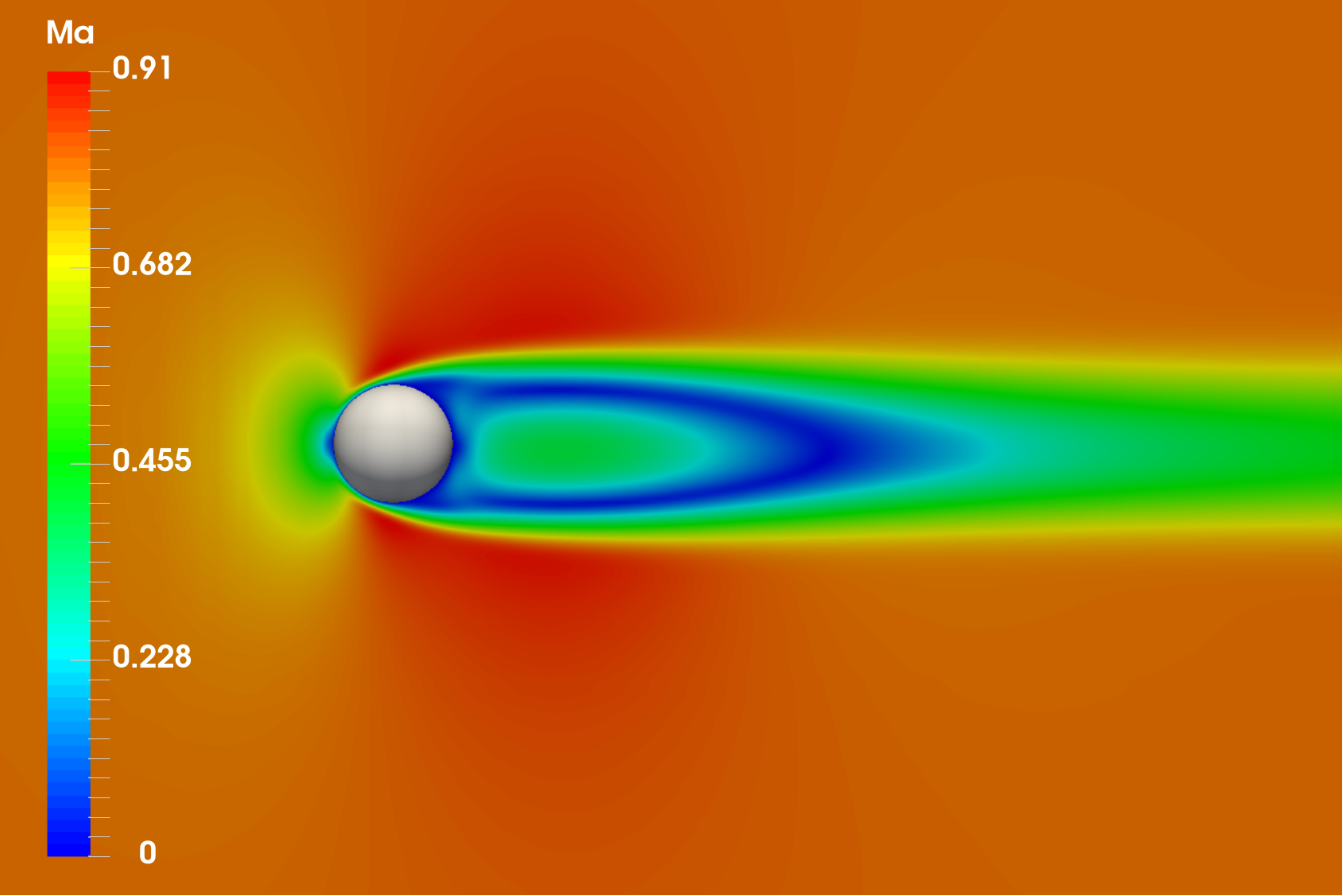} & \includegraphics[width=0.48\linewidth]{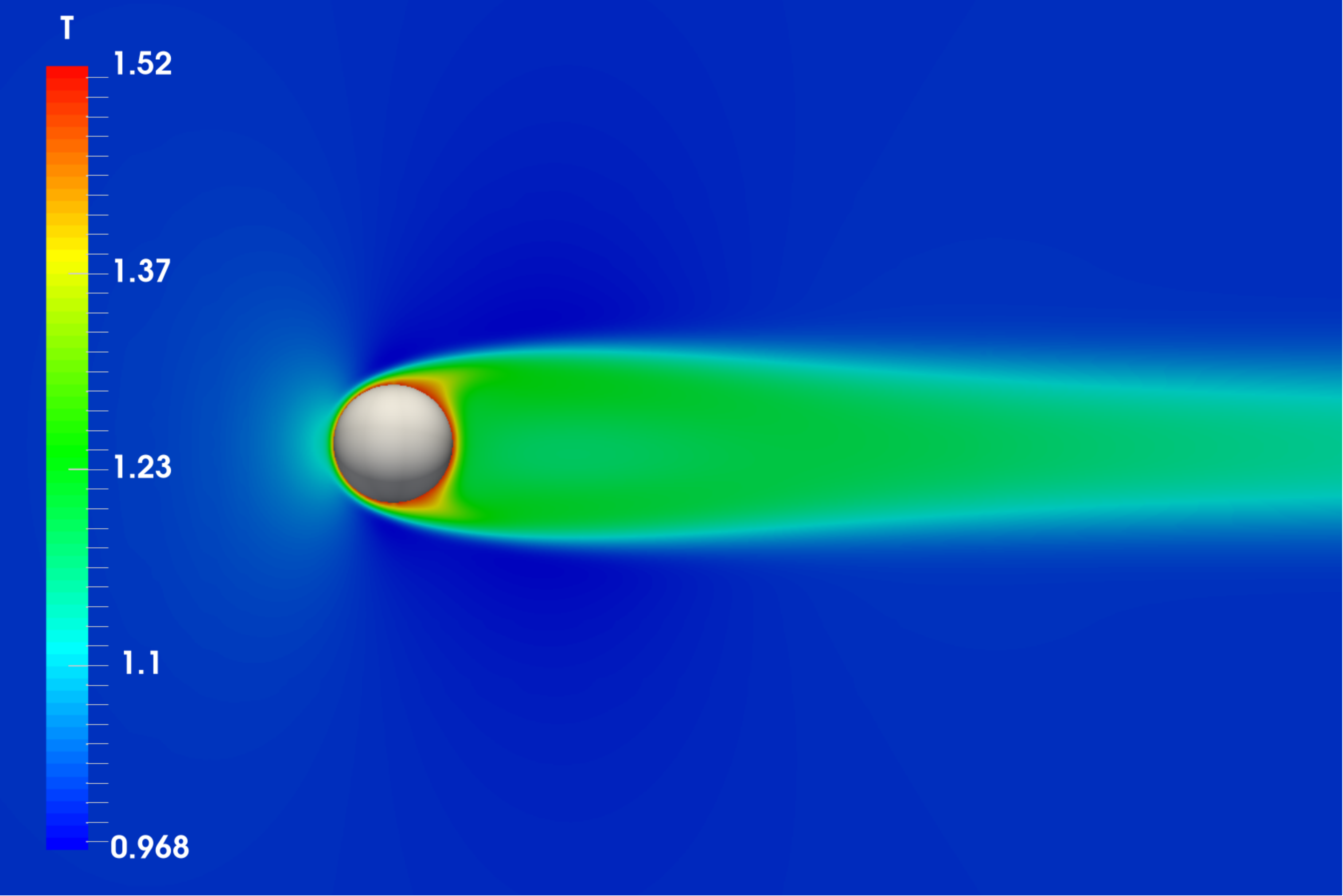} \\
(c) & (d) \\
\includegraphics[width=0.48\linewidth]{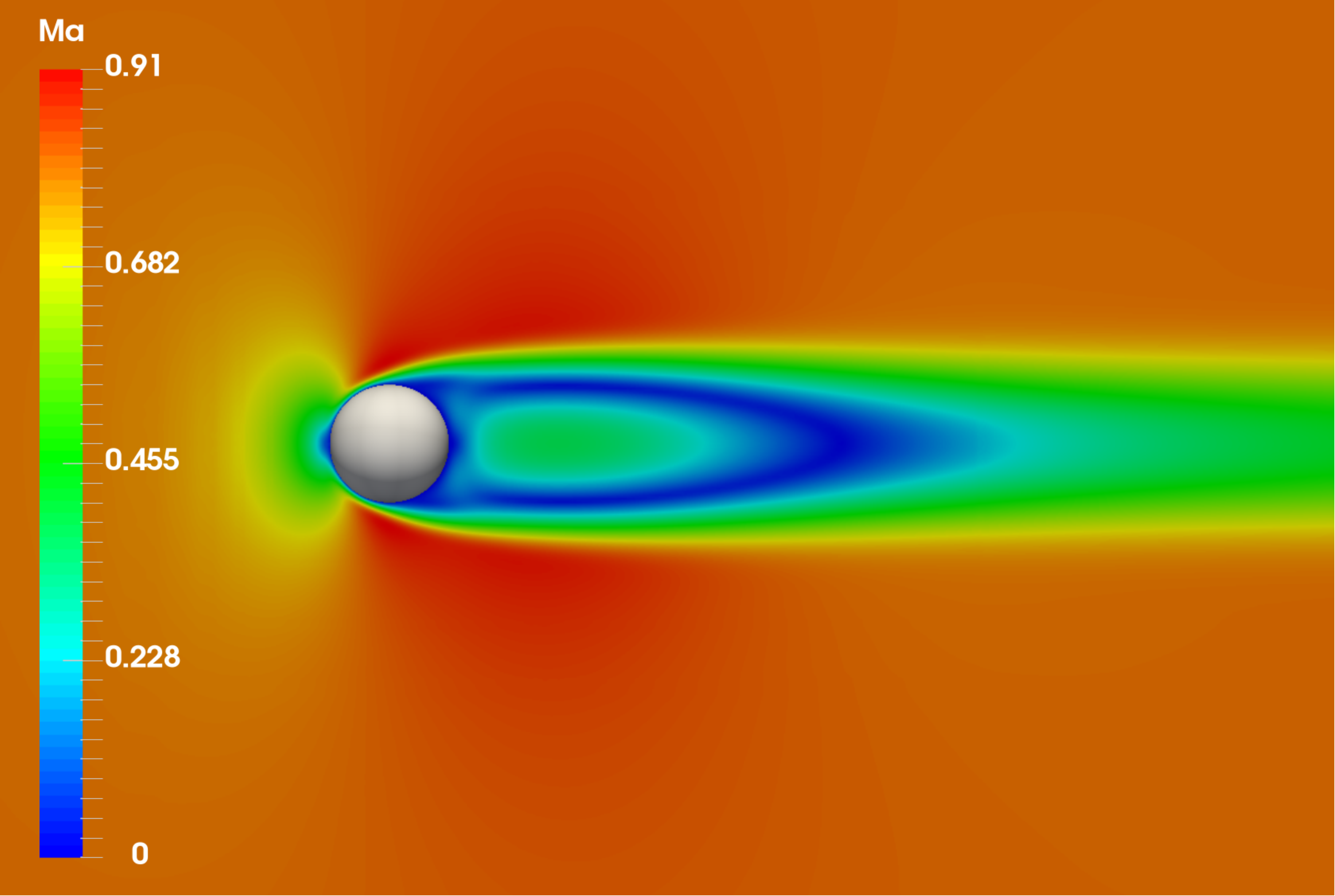} & \includegraphics[width=0.48\linewidth]{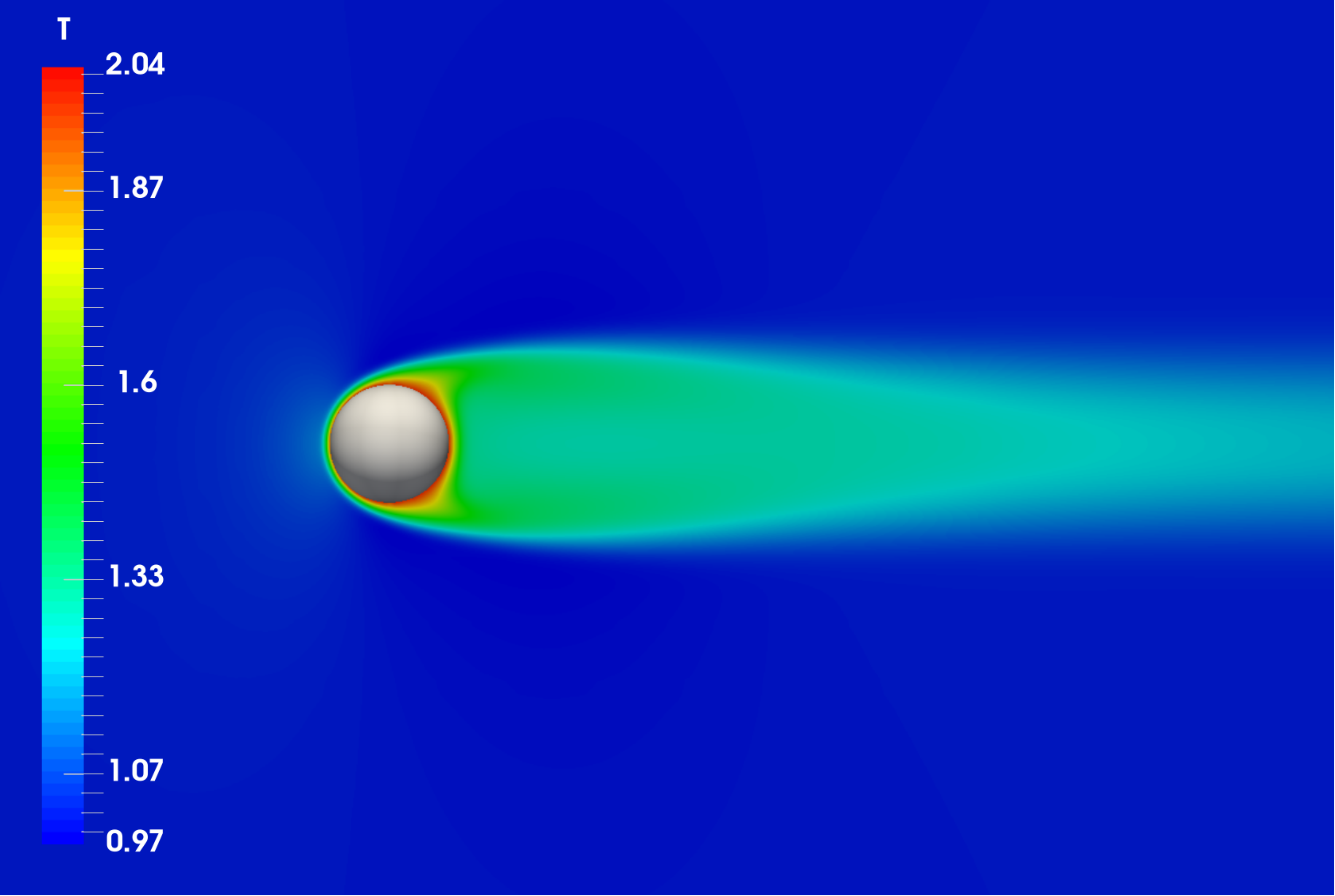} \\
(e) & (f)
\end{tabular}
\caption{\label{fig:3d_sphere} Iso-contours of $Ma$ number and
temperature for $Ma=0.8$ and $Re=300$: (a) $r_\theta=1.1$;(b)
$r_\theta=1.5$;(c) $r_\theta=2$}
\end{figure}

This test case has been chosen because of the emergence of different
regimes which exhibit a very high sensitivity to the values of
temperature ratio $r_\theta=1.1, \, 1.5, \, 2$, representing a challenging test case
of validation. We choose to investigate heated flow around sphere at
$Ma=0.8$ and $Re=300$ and the isocontours of the velocity field (represented in terms of $Ma$) and the temperature are shown in Figure
\ref{fig:3d_sphere} for the three cases. One can see that the flow undergoes a transition from an unsteady regime to a steady axisymmetric state. This effect is clearly driven by the heat exchange dynamics. Results are compared with
recent data reported in the literature for body-fitted numerical
simulations using high order discretization schemes \cite{Nagata_thermo} (Table \ref{ta:table6}). The physical quantities compared are the mean drag coefficient $\overline{C_d}$, the length of the recirculation bubble $l_s$ and the Strouhal number $St$. One can see that a reasonably good agreement is obtained, and in particular variations with different $r_{\theta}$ values are very similar. It appears that the IBM tends to provide a slight over prediction (around $10\%$) of the mean drag coefficient $\overline{C_d}$ when compared with body-fitted simulations. This result, which was already observed in previous studies \cite{Riahi_2018}, is probably associated with mesh resolution and the numerical schemes employed for the analysis.

\begin{table}[h!]
\centering
\begin{tabular}{c|c|c|c|c}
&Studies & $\overline{C_d}$  &  $l_s$ & $St$  \\ \hline
\textbf{Ma=0.8} &\textbf{Present ($r_\theta=1.1$)}  & 0.895 & 2.9 & 0.104  \\
\textbf{Re=300}&Nagata \cite{Nagata_thermo} (Num.)  & 0.81       & 2.8 & 0.105  \\\hline
&\textbf{Present ($r_\theta=1.5$)}  & 0.93       &3.18 &   \\
&Nagata \cite{Nagata_thermo} (Num.)  &    0.85    & 3.1 &   \\\hline
&\textbf{Present ($r_\theta=2$)}  & 0.99 & 3.4 &   \\
&Nagata \cite{Nagata_thermo} (Num.) &    0.91    & 3.35 &   \\\hline
\end{tabular}
\caption{Bulk flow quantities for the flow past a sphere obtained
via IBM simulation. Present results are compared with available data
in the literature. $\overline{C_d}$ is the mean, time-averaged drag coefficient,
$l_s$ is the recirculation length and $St$ is the Strouhal number.}
\label{ta:table6}
\end{table}

\subsection{Investigation of the supersonic flow around a sphere}
\begin{figure}
\begin{tabular}{cc}
\includegraphics[width=0.48\linewidth]{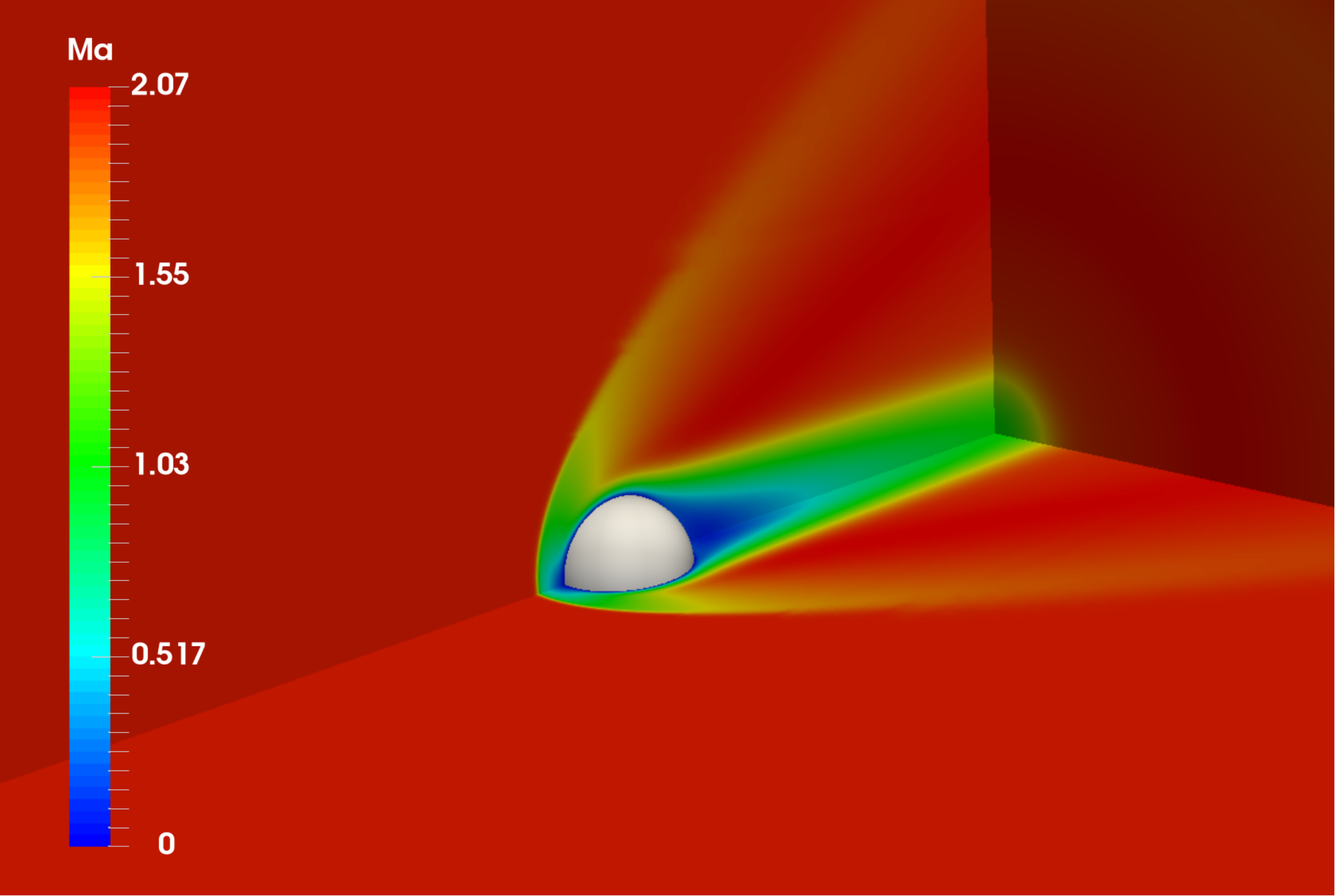} & \includegraphics[width=0.48\linewidth]{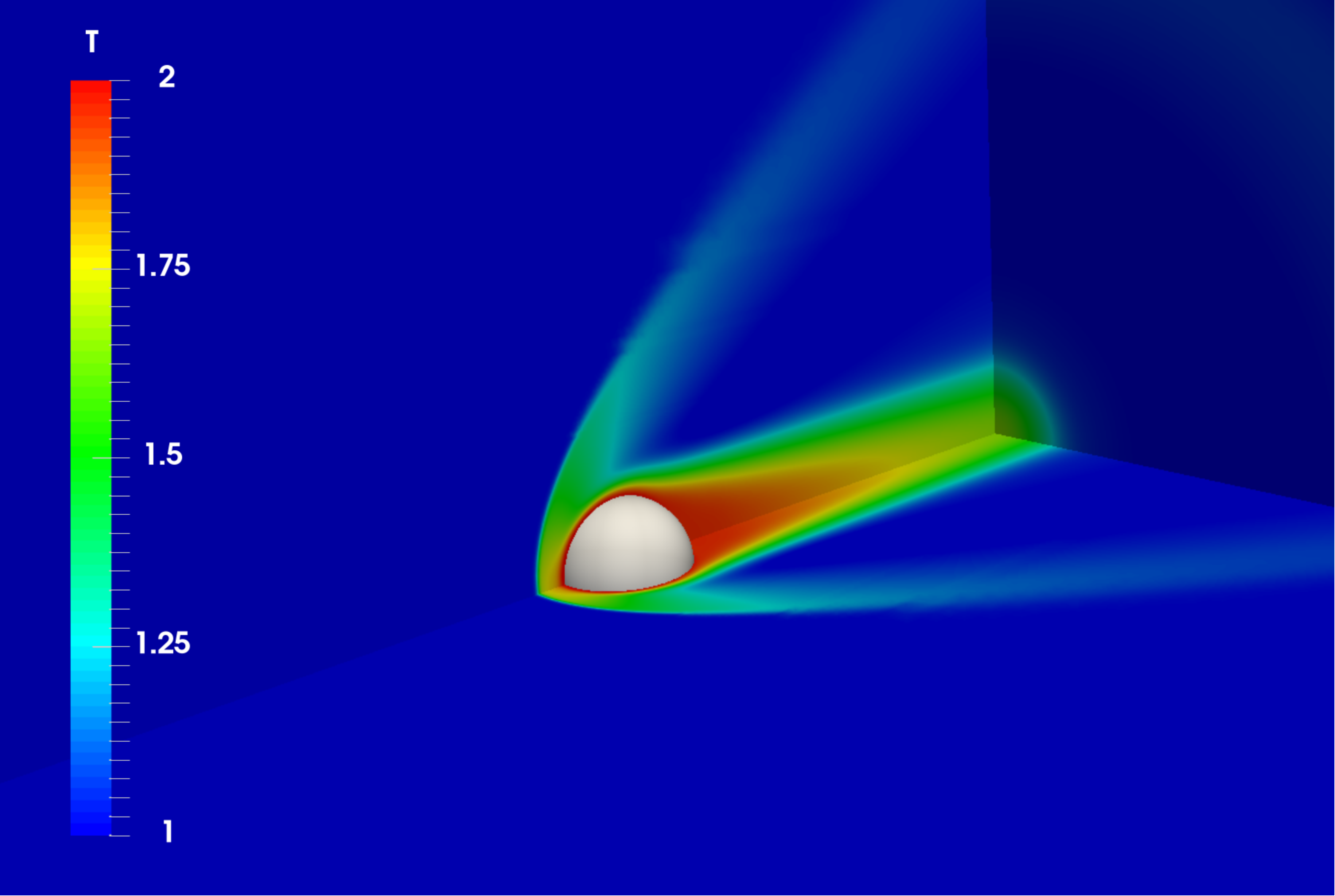} \\
(a) & (b)
\end{tabular}
\caption{\label{fig:superHighRe} Numerical simulation of the
supersonic flow around a sphere for $Ma=2$, $Re=300$ and
$r_\theta=2$ : (a) Isocontours of the $Ma$ number  (b) isocontours
of the Temperature}
\end{figure}

The supersonic flow for $Ma=2$, $Re=300$ and $r_{\theta}=2$ is now investigated, using the same mesh previously employed. In this case compressibility effects are
very strong and a steady axisymmetric configuration is observed, as qualitatively shown in figure \ref{fig:superHighRe}. The main bulk flow quantities are again compared with results obtained from body-fitted numerical simulations \cite{Nagata_thermo} and reported in table \ref{ta:table7}. It appears that all the physical
features are in agreement with the data in the literature \cite{Nagata_thermo}, assessing the potential of the IBM method developed. A slight over prediction  {(this time around $5 \%$) } of the drag coefficient $C_D$ is again observed, while the recirculation length $l_s$ and the distance of the shock from the wall $\Delta_{shock}$ are very similar. 

\begin{table}[h!]
\centering
\begin{tabular}{c|c|c|c|c}
&Studies & $\overline{C_d}$  &  $l_s$ & $\Delta_{shock}$ \\ \hline
\textbf{Ma=2}&\textbf{Present ($r_\theta=2$)}  & 1.48       & 0.85    & 0.23\\
\textbf{Re=300}&Nagata \cite{Nagata_thermo} (Num.) & 1.41       & 0.85  & 0.25 \\\hline
\end{tabular}
\caption{Bulk flow quantities for the flow past a sphere obtained
via IBM simulation. Present results are compared with available data
in the literature. $C_d$ is the time-averaged drag coefficient,
$x_s$ is the recirculation length, $Nu$ Nusselt number and
$D_{shock}$is  the shock distance.} \label{ta:table7}
\end{table}

\section{Fluid - Structure interaction : oscillating cylinder with fixed wall temperature}\label{sec:FSI}

In this last section, the flow around an oscillating two-dimensional
cylinder is investigated. The simulation is performed using the $
IBM-HT-rhoCentralFoam $ solver.

\begin{figure}
\begin{tabular}{cc}
\includegraphics[width=0.48\linewidth]{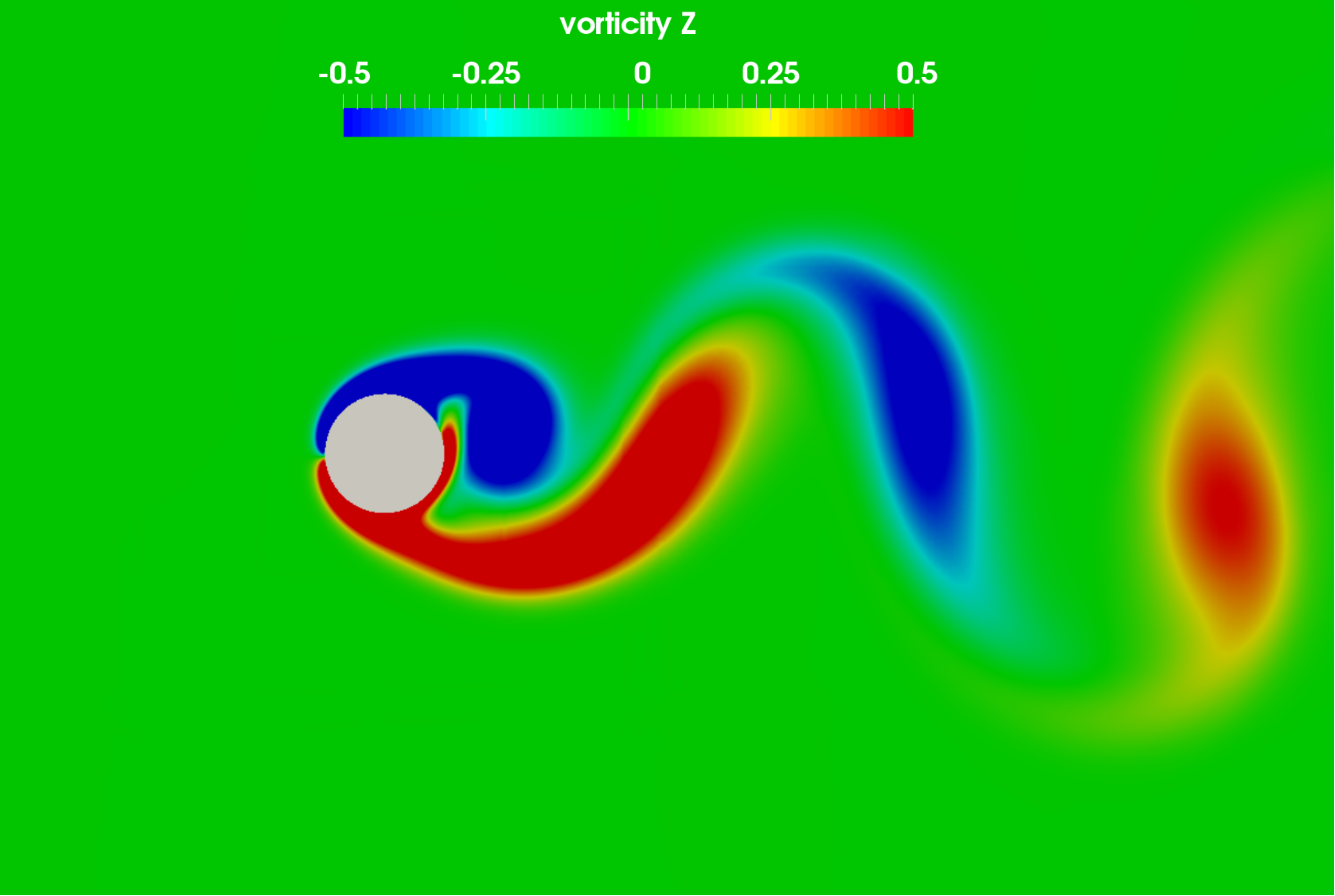} & \includegraphics[width=0.48\linewidth]{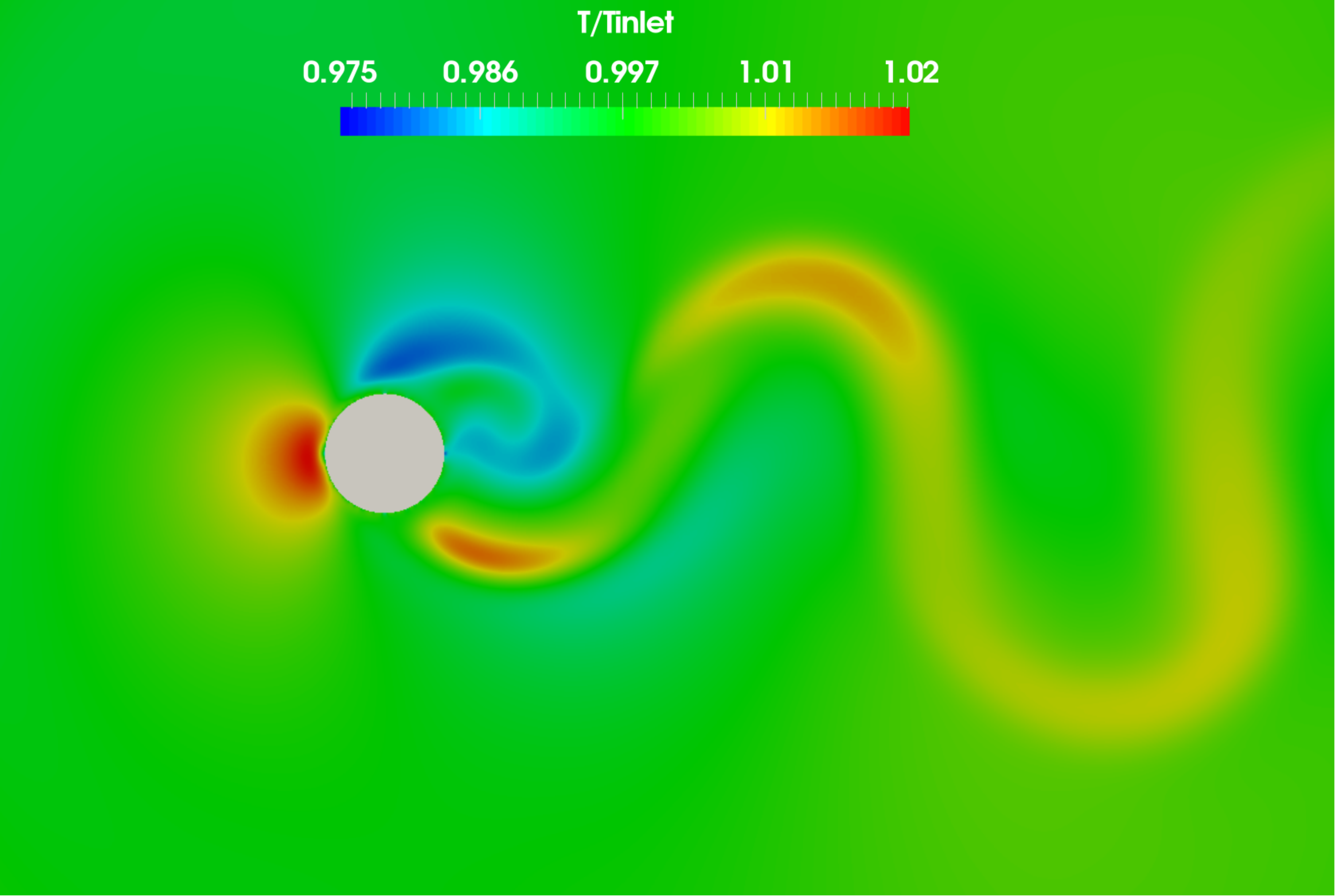} \\
(a) & (b)
\end{tabular}
\caption{\label{fig:mvt_thermique} Iso-contours vorticity (a) and
temperature (b) for the flow around a longitudinal oscillating
cylinder $Re=100$, $r_\theta=1$ }
\end{figure}

\begin{figure}
\begin{tabular}{cc}
\includegraphics[width=0.48\linewidth]{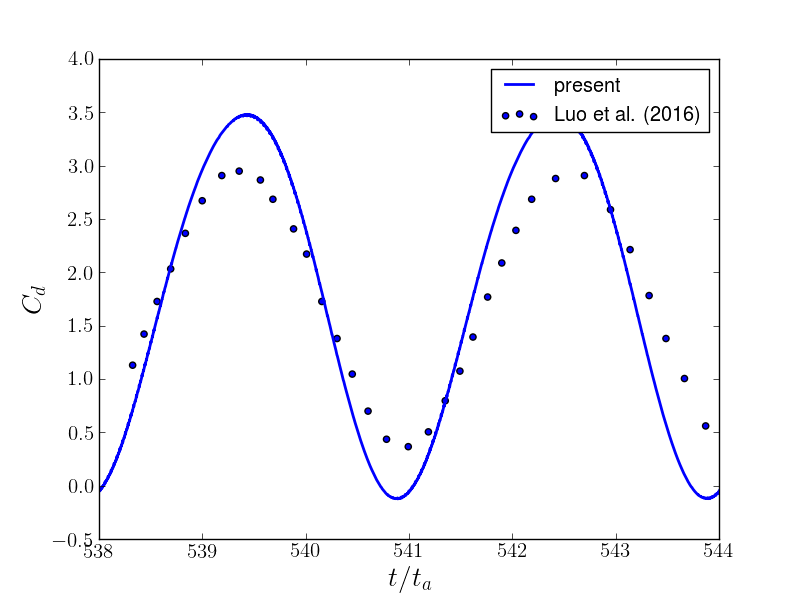} & \includegraphics[width=0.48\linewidth]{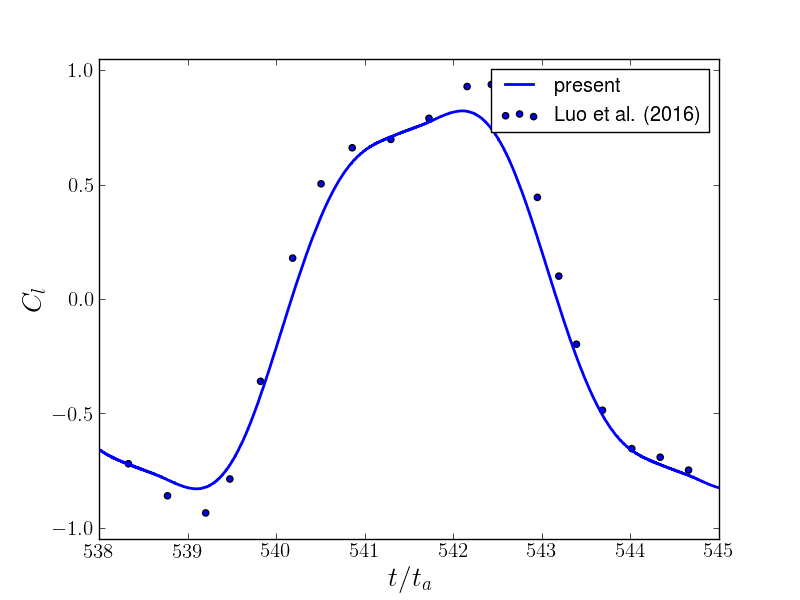} \\
(a) & (b)
\end{tabular}
\caption{\label{fig::cd_cl_osc_therm_com}  {Comparison of the (a) 
drag coefficient $C_d$ and (b) lift coefficient $C_l$ for the flow around a longitudinal oscillating cylinder, $r_\theta=1$ with results by Luo et al. \cite{Luo_2016}}}
\end{figure}

The motion of the cylinder is imposed in the streamwise direction $x$, providing an explicit law of movement for the Lagrangian markers following a
sinus law. More precisely, the position in the streamwise direction of the Lagrangian
markers is updated in time following the equation:
\begin{equation}
X_s (t) = X_s (0) + A sin (2 \pi f_c t)
\end{equation}

with $A=0.14 \times D$ and $f_c=2 \times f_0$. A preliminary
calculation is performed on a fixed cylinder to determine the
shedding frequency $f_0$. The values of the parameters have been
chosen in order to allow comparison with a study proposed by Luo et
al. \cite{Luo_2016} with  {$Ma=0.3$}, $Re=100$ and $r_\theta=1$. In figure
\ref{fig:mvt_thermique} the qualitative evolution of the isocontours
of the temperature and of the vorticity are shown after the initial
transient. The comparison of present results with the findings by
Luo et al. \cite{Luo_2016} it terms of the time evolution
of  { the drag coefficient $C_d$} and lift coefficient $C_l$ is very good, as shown in figure \ref{fig::cd_cl_osc_therm_com}. Therefore, the \textit{IBM-HT} model has proven good accuracy even for a more complex case including fluid-structure interaction.

\begin{figure}
\begin{tabular}{cc}
\includegraphics[width=0.48\linewidth]{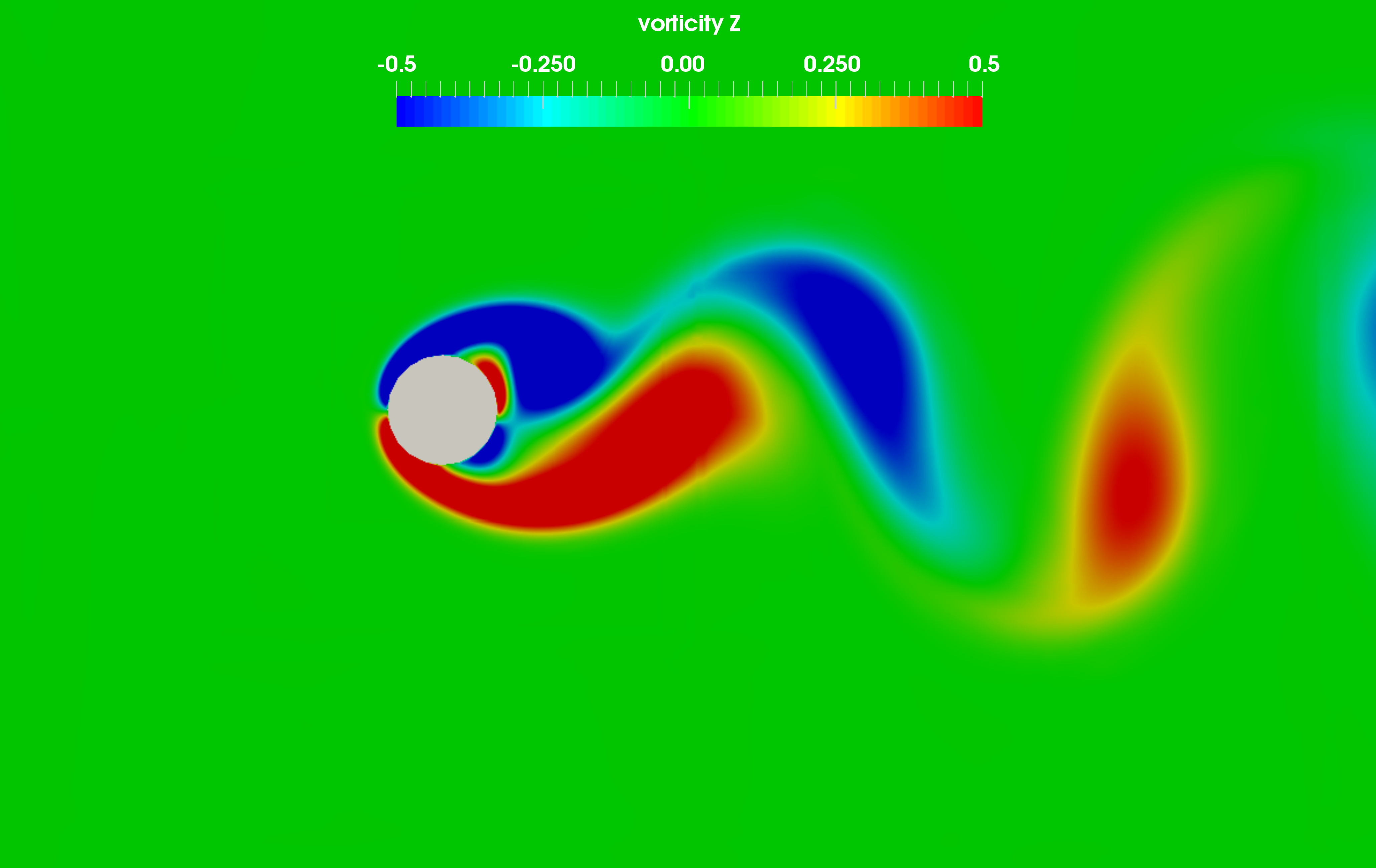} & \includegraphics[width=0.48\linewidth]{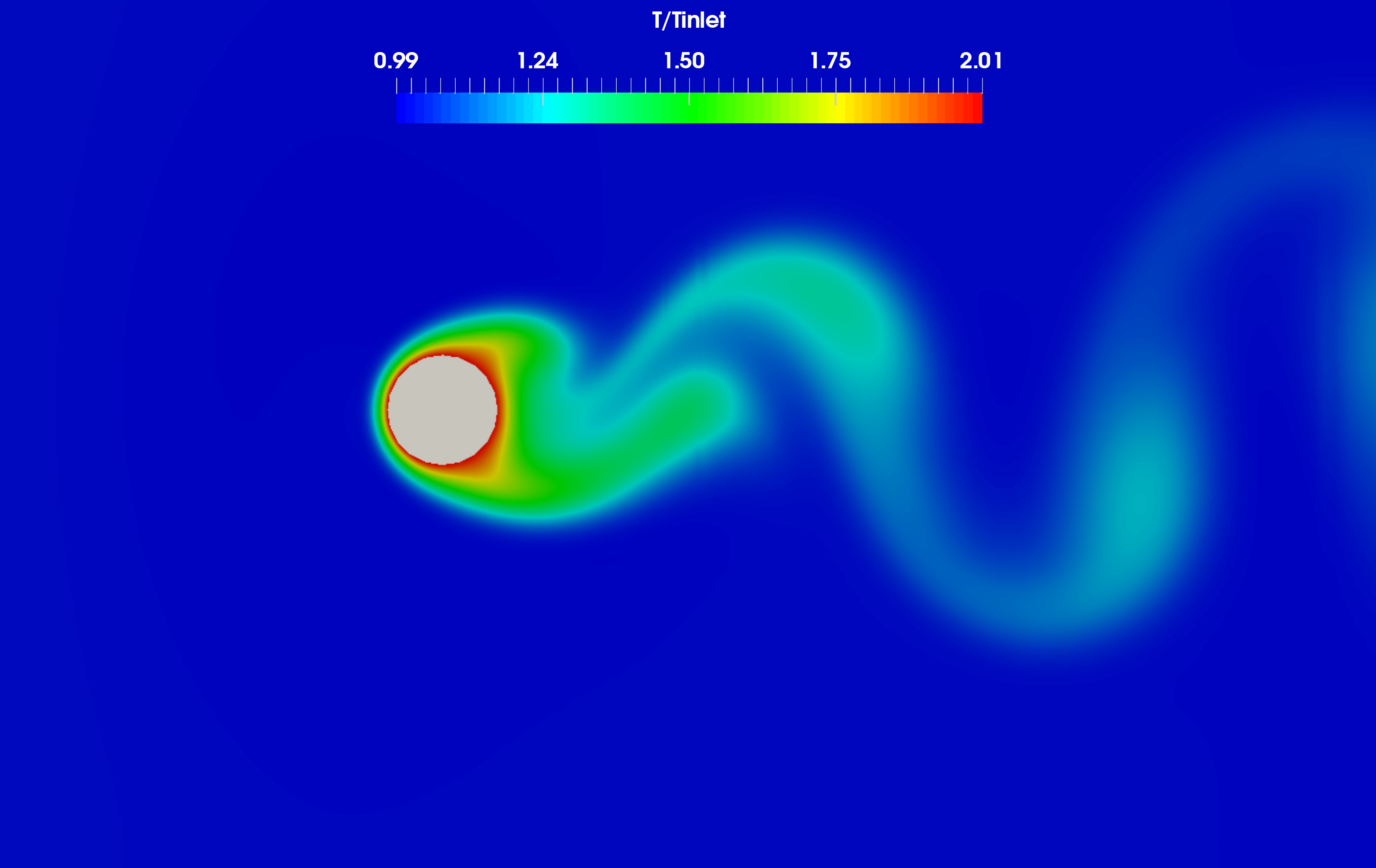} \\
(a) & (b)
\end{tabular}
\caption{\label{fig:mvt_thermique_t2}  {Iso-contours vorticity (a) and
temperature (b) for a flow around a longitudinal oscillating
cylinder with $r_{\theta}=2$ }}
\end{figure}

\begin{figure}
\begin{tabular}{cc}
\includegraphics[width=0.48\linewidth]{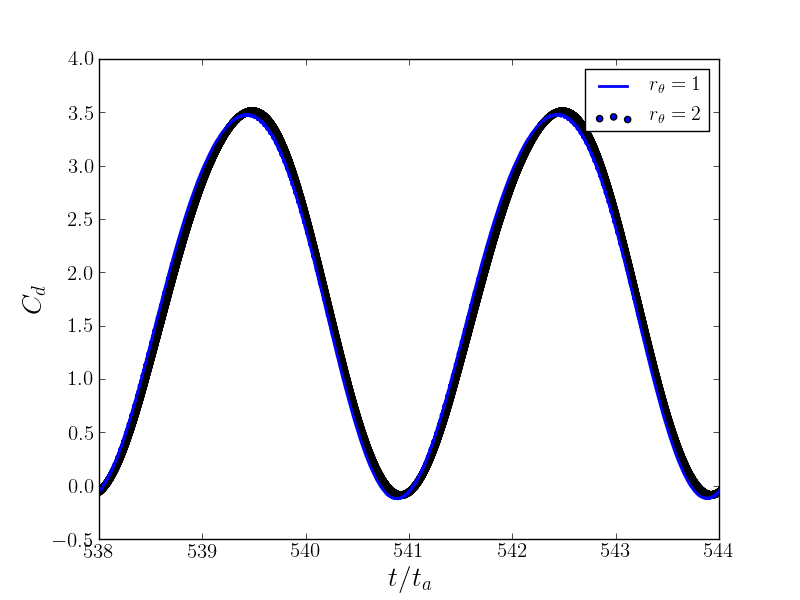} & \includegraphics[width=0.48\linewidth]{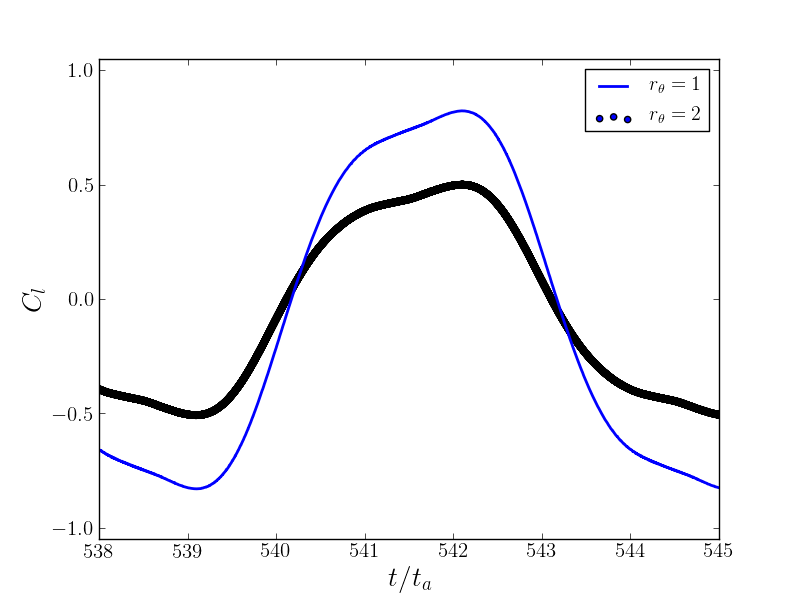} \\
(a) & (b)
\end{tabular}
\caption{\label{fig::cd_cl_osc_therm_t_2} { (a)  Temporal evolution of the
drag coefficient $C_d$ and (b) of the
lift coefficient $C_l$ for flow around an oscillating circular
cylinder. The cases for $r_{\theta}=1$ and $r_{\theta}=2$ are compared}}
\end{figure}

 {An additional simulation with imposed temperature $r_{\theta}=2$ was carried out in order to study the effects of the increase in temperature on the aerodynamic parameters \ref{fig:mvt_thermique_t2}. The variation in temperature affects the viscosity close to the wall, which modifies the aerodynamic performance.}  {The comparison of the drag coefficient $C_d$ and lift coefficient $C_l$ for the two cases, which are shown in fig. \ref{fig::cd_cl_osc_therm_t_2}, shows important differences. In particular, the maximum value of $C_l$ decreases for the case with $r_\theta=2$, in agreement with the findings by Homsi et al. \cite{Homsi_2021}.}

\section{Conclusion}\label{sec:conclusion}
In this paper, a discrete-forcing immersed boundary method was
proposed for simulating both heat exchange and compressible flows, for
stationary and moving bodies. The proposed methodology introduces an
additional heat source term to the right hand side of the
energy equation to keep the solid boundary at the prescribed
temperature. The compressible Navier-Stokes equations are
discretized on a Cartesian grid and solved
by either the sonicFoam or the rhoCentralFoam solver of the OpenFOAM
platform according to the Mach number of the flow.

To validate the robustness and accuracy of the proposed method, a
series of simulations was  performed on canonical forced convection
flows with comparison with existing results. Both the flow dynamics
and the heat transfer were in good agreement with the data in
literature. The 3D flows around an isothermal sphere in the
transonic and supersonic regimes were investigated. Different wall
temperature conditions were applied leading to the transition
between an unsteady periodical wake and a steady axisymmetric state.
The results obtained with the proposed method were found in close
agreement with recent direct numerical simulations available in the
literature, demonstrating the capability of the numerical strategy
to treat both compressible and heated flows with satisfactory
accuracy. Finally the method was successfully validated on a
benchmark case involving moving boundaries, for which the
computational cost of remeshing techniques would be significantly high for classical body-fitted approaches. 
Future works envision an improvement in accuracy for higher Reynolds number configurations, including wall modelling within the formulation of the IBM model.

\section*{Acknowledgement}

{The research work has been developed using computational resources within the framework of the project gen7590-A0012A07590 DARI-GENCI.}

\bibliographystyle{unsrt}
\bibliography{allbiblio}

\appendix

\section{IBM-HT-sonicFoam}
\label{apendixA}

The complete algorithm with integration of the IBM method can be summarized in the following steps:

\begin{enumerate}

\item{The discretized continuity, momentum and energy equations are resolved,
providing a first time advancement of $\rho^{\star}$,
$\mathbf{u}^{\star}$ and $\theta^{\star}$. These quantities are calculated imposing $\mathbf{f}=0$ and $q=0$.}
\item{The fields calculated in step $1$ are \textit{interpolated} on the
Lagrangian markers in order to obtain the value of the forcing
$\mathbf{F}$ and the source term $Q$. These quantities are
\textit{spread} over the Eulerian mesh, in order to provide the
contributions for $\mathbf{f}$ and $q$ in each Eulerian mesh
element. }
\item{The whole system is resolved again, starting from stored quantities for the
time step $n$ and but now including the source terms previously
calculated for the momentum equation and the internal energy
equation. An iterative procedure is triggered until convergence:
\begin{eqnarray}
\label{eq:rho_sonicF}
\rho^{n+1} &=& \frac{\phi_{\rho} (\rho^{\star}, \mathbf{u}^{\star})}{a_{\rho}} \\
\label{eq:u_sonicF}
\mathbf{u}^{n+1} &=& \frac{\phi_{\mathbf{u}} (\rho^{n+1}, \mathbf{u}^{\star})}{a_\mathbf{u}} - \frac{\mathbf{grad} p^{\star}}{a_\mathbf{u}} + \frac{\mathbf{f}}{a_\mathbf{u}} \\
\label{eq:e_sonicF}
e^{n+1} &=& \frac{\phi_{e} (\rho^{n+1}, \mathbf{u}^{n+1}, e^{\star})}{a_e} -\frac{div (p^{\star} \mathbf{u}^{n+1})}{a_e} + \frac{q}{a_e}\\
\label{eq:p_sonicF}
p^{n+1} &=& \frac{\phi_p (p^{\star},\rho^{n+1},\mathbf{u^{n+1}})}{a_p} + \frac{\phi_{fp}(\mathbf{f})}{a_p}
\end{eqnarray}
}
\end{enumerate}

In this case, the terms $\mathbf{f}$ and $q$ are not updated during
the step 3. They are calculated only once in step 2.

\section{IBM-HT-rhoCentralFoam}
\label{apendixB}

The integration of the \textit{IBM-HT} method in the solver
rhoCentralFoam presented in \cite{Riahi_2018} follows these steps:
\begin{enumerate}
\item {A prediction step resolving continuity, momentum and energy equations is performed
in order to obtain first estimations for $\rho^{\star}$ ,
$e^{\star}$ and $\mathbf{u}^{\star}$ (and $p^{\star}$ via an
equation of state). The volume sources are here
$\mathbf{f}=\mathbf{0}$ and $q=0$.}
\item The physical quantities $\rho^{\star}$ , $p^{\star}$, $e^{\star}$ and $\mathbf{u}^{\star}$ are
interpolated in the Lagrangian space and $\mathbf{F}$ and $Q$ are
calculated. This field is \textit{spread} over to the Eulerian mesh,
so that the value of the forcing terms $\mathbf{f}$ and $q$ for each
mesh cell is calculated.
\item Equations of the first step are resolved again including the IBM forcing:
\begin{eqnarray}
\label{eq:rho_centralF}
\rho^{n+1} &=& \frac{\phi_{\rho} (\rho^{\star}, \mathbf{u}^{\star})}{a_{\rho}} \\
\label{eq:u_centralF1}
\mathbf{(\rho u)}^{\star \star} &=& \frac{\phi^{\prime}_{\mathbf{u}} ((\rho \mathbf{u})^{\star})}{a_\mathbf{u}} - \frac{\mathbf{grad} p^{\star}}{a_\mathbf{u}} \\
\label{eq:u_update}
\mathbf{(u)}^{\star \star} &=& \mathbf{(\rho u)}^{\star \star}/ \rho^{n+1} \\
\label{eq:u_centralF2}
\rho ^{n+1} \mathbf{ u}^{n+1} &=& \rho ^{n+1} \mathbf{ u}^{\star \star} + \frac{\phi_{\mathbf{u}} (\rho^{\star}, \mathbf{u}^{\star})}{a_\mathbf{u}} - \frac{\phi^{\prime}_{\mathbf{u}} ((\rho \mathbf{u})^{\star})}{a_\mathbf{u}} + \frac{\mathbf{f}}{a_\mathbf{u}}  \\
\label{eq:e_centralF1}
(\rho {e_t})^{\star \star} &=& \frac{\phi^{\prime}_{e_t} ((\rho {e_t})^{\star}, \mathbf{u}^{\star \star})}{a_{e_t}} -\frac{div (p^{\star} \mathbf{u}^{\star})}{a_{e_t}} \\
\label{eq:e_update}
{e}^{\star \star} &=& (\rho {e_t})^{\star \star} / \rho^{n+1}-0.5(\mathbf{(u)}^{\star \star}.\mathbf{(u)}^{\star \star}) \\
\label{eq:temperature_update}
{\theta}^{\star \star} &=& {e}^{\star \star} / c_v \\
\label{eq:e_centralF2}
\rho ^{n+1} e^{n+1} &=& \rho ^{n+1} e^{\star \star}+ \frac{\phi_{e} (\rho^{n+1}, \mathbf{u}^{n+1}, e^n)}{a_e} - \dfrac{div(\lambda(\theta^{\star \star}) \mathbf{grad}(\theta^{\star \star}))}{a_e} \nonumber\\&-&\frac{\phi^{\prime}_{e_t} ((\rho {e_t})^{\star}, \mathbf{u}^{\star \star})}{a_{e_t}}  + \frac{q}{a_e}
\end{eqnarray}
\item Finally, the temperature $\theta^{n+1}=e^{n+1}/C_v$ and the pressure $p^{n+1}=\rho^{n+1} \cdot (r \theta^{n+1})$ are updated.
\end{enumerate}

\section{Grid convergence analysis}
\label{sec:appC} 

{
The accuracy of the proposed IBM method is assessed via the analysis of the flow around a circular cylinder for $Ma=2.5$ and $Re=250$. No heat exchange in considered in this validation.

A grid convergence analysis is performed evaluating results using four different grids. The mesh resolution in the near cylinder region is imposed to be $\Delta x = \Delta y = \{ \dfrac{D}{10} ; \dfrac{D}{15} ; \dfrac{D}{25} ; \dfrac{D}{50} \}$ where $D$ is the diameter of the cylinder. The corresponding number of Lagrangian markers employed is $\{ 28 ; 39 ; 62 ; 104 \}$, respectively.} Data from the most refined mesh is used as a reference solution. 

The precision of the IBM method is investigated using $L_\infty$ norms so that, for a physical quantity $\phi$, the error is estimated as:
{
\begin{eqnarray}
e_{\phi_{L_\infty}} = \parallel \phi_{ref} - \phi_G \parallel_\infty
\end{eqnarray}
}
{ where $\phi_G$ is the reference solution. }

\begin{figure}[ht]
\centering
\includegraphics [width=0.75\linewidth]{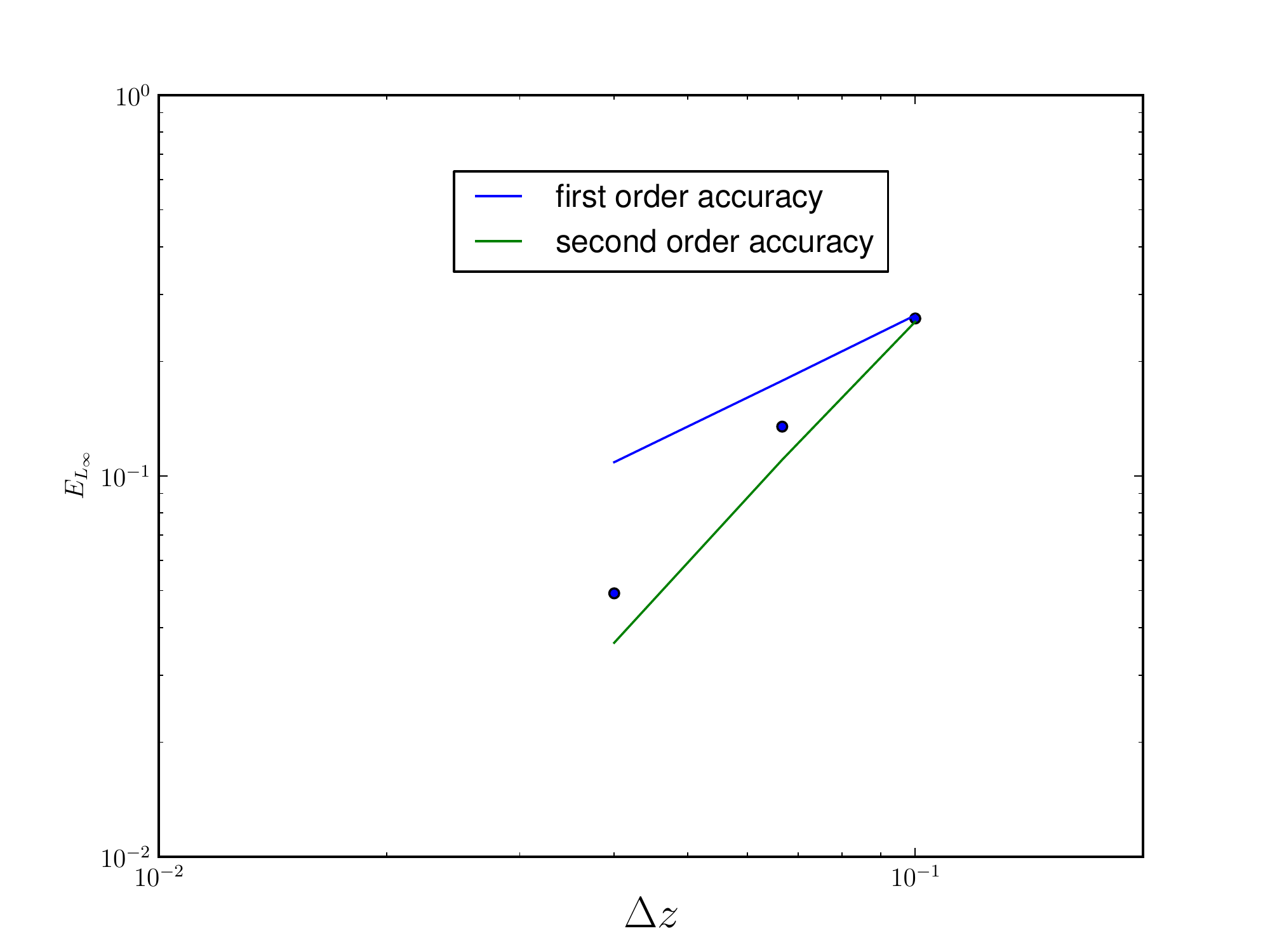}
\caption{\label{fig::error_Fibm} {Convergence rate in the prediction of the drag coefficient $C_D$ using the IBM method presented in this work.}}
\end{figure}

{
The behavior of error in the prediction of the drag coefficient $C_D$ is shown in Figure \ref{fig::error_Fibm}. In the framework of this IBM method, the drag coefficient is directly calculated using information available on the Lagrangian markers. For this quantity, the rate of convergence is almost $2$.}

\begin{figure}
\begin{tabular}{cc}
\includegraphics[width=0.48\linewidth]{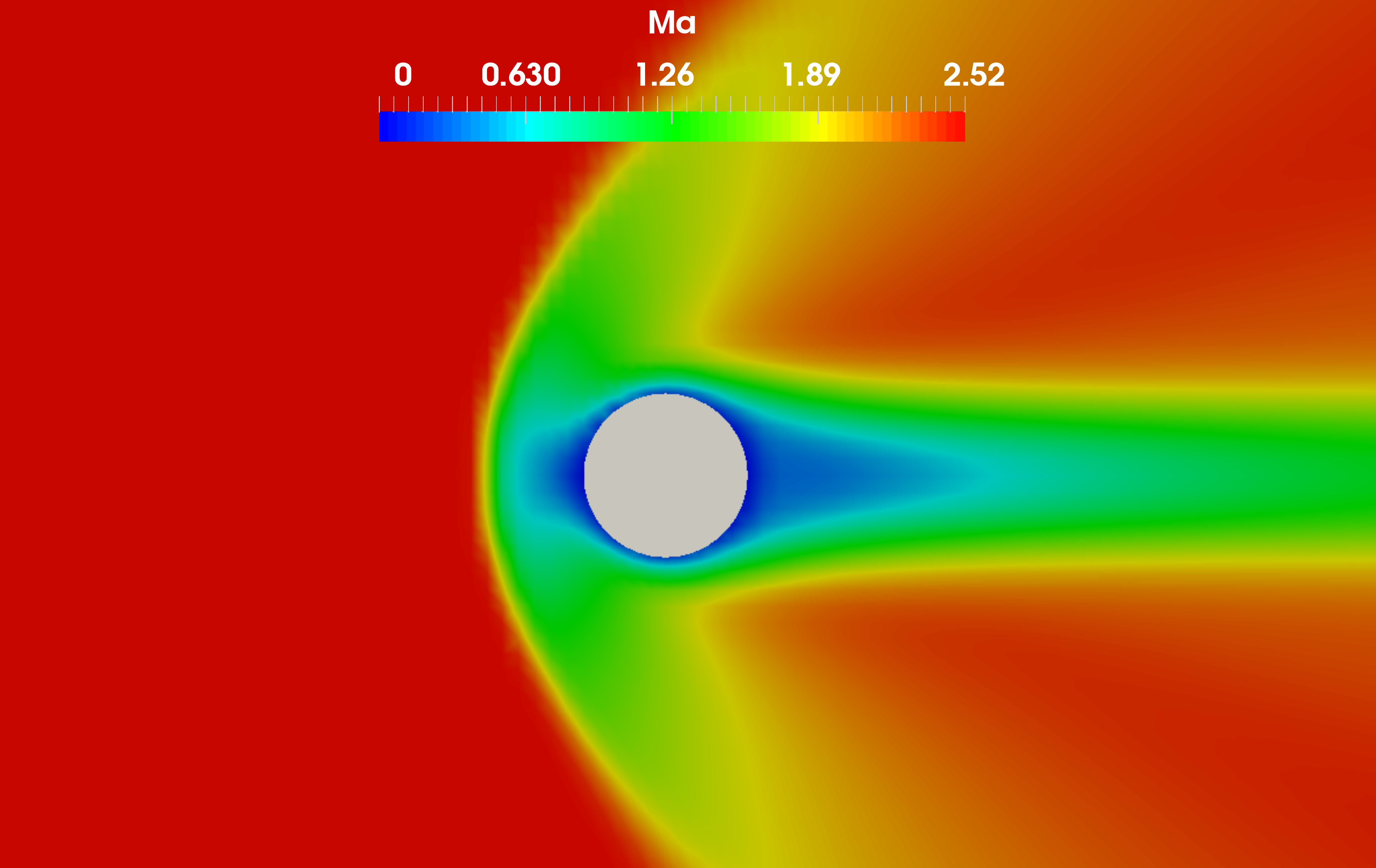} & \includegraphics[width=0.48\linewidth]{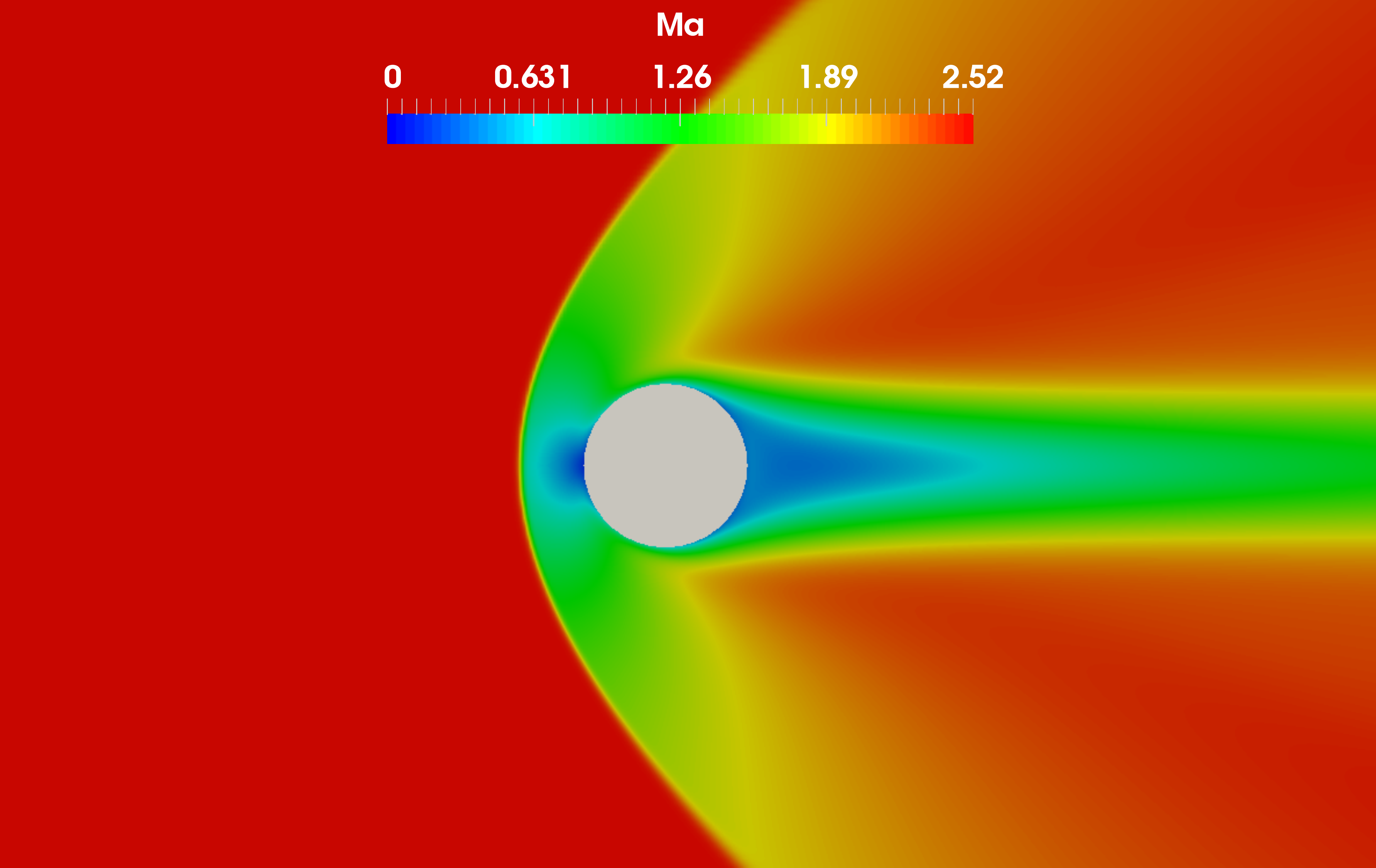} \\
(a) & (b) 
\end{tabular}
\caption{\label{fig::ma_2_5} Isocontours of $Ma$ number : (a) $\Delta x = \Delta y = 0.1$ and (b)  $\Delta x = \Delta y = 0.02$}

\end{figure}

\begin{figure}
\centering \includegraphics[width=0.8\linewidth]{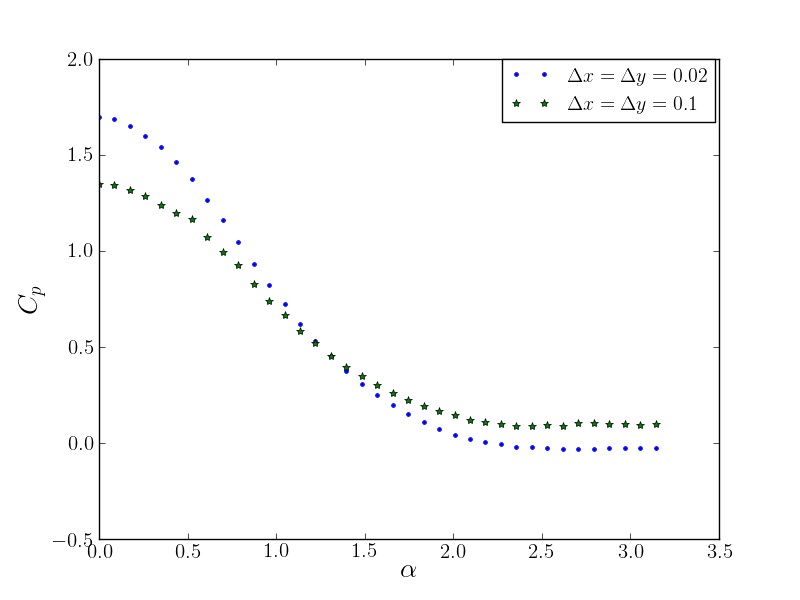}
\caption{\label{fig::ma_2_5_cp} Pressure coefficient distributions for the cases $\Delta x = \Delta y = 0.1$ and $\Delta x = \Delta y = 0.02$}
\end{figure}

In addition, the qualitative evolution of the isocontours of the Mach number is presented in fig. \ref{fig::ma_2_5} for the coarsest and the finest mesh resolution. Despite the main flow features are captured by both simulations, one can see differences in the precision in particular close to the shock region, which is expected.
The distribution of pressure coefficient $C_p = 2 (p - p_{\infty}) / (\rho_{\infty} U_{\infty}^2)$ around the cylinder, which are shown in fig. \ref{fig::ma_2_5_cp} against the azimuthal angle $\alpha$, allow to draw the same conclusions. The qualitative behavior of the coarsest and finest simulation is very similar, but differences can be observed in particular for the stagnation point for $\alpha=0$. 

\end{document}